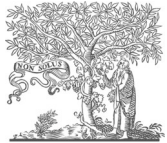
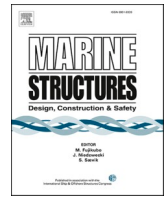

# Marginal and total exceedance probabilities of environmental contours

Ed Mackay [a,*], Andreas F. Haselsteiner [b,c]

[a] *College of Engineering, Mathematics and Physical Sciences, University of Exeter, Penryn, UK*
[b] *University of Bremen, Institute for Integrated Product Development (BIK), Bremen, Germany*
[c] *ForWind – Center for Wind Energy Research of the Universities of Oldenburg, Hannover and Bremen, Germany*

A B S T R A C T

Various methods have been proposed for defining an environmental contour, based on different concepts of exceedance probability. In the inverse first-order reliability method (IFORM) and the direct sampling (DS) method, contours are defined in terms of exceedances within a region bounded by a hyperplane in either standard normal space or the original parameter space, corresponding to marginal exceedance probabilities under rotations of the coordinate system. In contrast, the more recent inverse second-order reliability method (ISORM) and highest density (HD) contours are defined in terms of an isodensity contour of the joint density function in either standard normal space or the original parameter space, where an exceedance is defined to be anywhere outside the contour. Contours defined in terms of the total probability outside the contour are significantly more conservative than contours defined in terms of marginal exceedance probabilities. In this work we study the relationship between the marginal exceedance probability of the maximum value of each variable along an environmental contour and the total probability outside the contour. The marginal exceedance probability of the contour maximum can be orders of magnitude lower than the total exceedance probability of the contour, with the differences increasing with the number of variables. For example, a 50-year ISORM contour for two variables at 3-h time steps, passes through points with marginal return periods of 635 years, and the marginal return periods increase to 10,950 years for contours of four variables. It is shown that the ratios of marginal to total exceedance probabilities for DS contours are similar to those for IFORM contours. However, the marginal exceedance probabilities of the maximum values of each variable along an HD contour are not in fixed relation to the contour exceedance probability, but depend on the shape of the joint density function. Examples are presented to illustrate the impact of the choice of contour on simple structural reliability problems for cases where the use of contours defined in terms of either marginal or total exceedance probabilities may be appropriate. The examples highlight that to choose an appropriate contour method, some understanding about the shape of a structure's failure surface is required.

## 1. Introduction

Extreme responses of marine structures are often calculated using the environmental contour method [1,2]. The environmental contour method provides a computationally efficient approximation to a full long-term analysis [3], sometimes referred to as the 'all sea states approach'. The method involves first estimating a model for the joint distribution of two or more environmental variables. The joint distribution is then used to calculate a set of points which have equal joint exceedance probability, that define a contour in two dimensions, a surface in three dimensions or a hypersurface in higher dimensions. The set of points is commonly referred to as a contour, regardless of the number of dimensions. The responses of the structure are then estimated at a discrete number of points along the contour, so-called design conditions, and the maximum response along the contour is compared with a maximum allowable response.

\* Corresponding author.
*E-mail address:* e.mackay@exeter.ac.uk (E. Mackay).

The objective of the environmental contour (EC) method is to find a region of the environmental parameter space (referred to here as the 'design region') such that a structure that can withstand all environmental conditions within that region has a probability of failure of less than or equal to a specified value. The environmental contour represents the boundary of the design region (see Fig. 1). The various methods used for constructing environmental contours make different assumptions about the shape of the failure region (the region of the environmental parameter space in which the structure fails) and consequently what constitutes an exceedance of the environmental contour.

The random nature of the structural response conditional on environmental parameters is not modelled explicitly in the EC method, and instead the structure is assumed to fail whenever the environmental conditions are within the failure region. To account for the effect of neglecting the short-term variability of the response, the contour exceedance probability is adjusted by an 'inflation factor' (see Ref. [4]). The maximum response along the $T$-year contour is then taken as an estimator of the $T$-year response.

Unlike the univariate case, there is no unique definition of multivariate exceedance (see e.g. Ref. [5,6]). Consequently, there are multiple ways to define an environmental contour. The commonly used Inverse First-Order Reliability Method (IFORM) [4] and direct sampling (DS) method [7] both make the assumption that the failure region is convex and that the failure surface can be approximated as a hyperplane. This assumption is equivalent to defining the contour exceedance probability as a marginal exceedance probability under a rotation of the axes.

When the failure region is convex, the FORM and DS methods give a conservative approximation to the failure probability (assuming that the contour exceedance probability has been adjusted to account for short-term variability of the response). When the failure region is concave, the IFORM and DS methods will underestimate the true failure probability. To address this issue, the Inverse Second-Order Reliability Method (ISORM) [8] and highest density (HD) region [9] environmental contour methods are both defined in terms of the total probability outside the contour. This results in a failure probability that is always conservative and requires no assumptions about the shape of the failure region.

The probability that an observation falls anywhere outside a contour is, by definition, larger than the probability an observation is larger than the upper bound of the contour in a particular dimension (the marginal exceedance probability). Therefore if the same exceedance probability is used to define both types of contour, then the ISORM and HD contours will be significantly more conservative than the IFORM or DS contours.

Comparisons between IFORM and DS contours have been presented in previous studies (e.g. Ref. [10–13]). However, the relationship between contours defined in terms of marginal exceedance probabilities and those defined in total exceedance probabilities has not previously been studied in detail, although example comparisons were presented in Refs. [8,9]. The purpose of this article is to provide a comparison of the relative levels of conservatism between contours defined in terms of marginal and total exceedance probabilities. In particular, the objective is to quantify the marginal exceedance probabilities associated with ISORM and HD contours and the total exceedance probabilities for IFORM and DS contours.

Understanding the total exceedance probabilities associated with IFORM and DS contours is useful for several purposes. Firstly, the total exceedance probability of the contour provides an upper bound on the failure probability of structure designed to withstand all conditions along the contour (under the simplifying assumption of a deterministic response). Secondly, knowing the total exceedance probability of a contour is useful when assessing the fit of a joint probability model to the data, as it can be used to quantify the expected number of observations outside the contour.

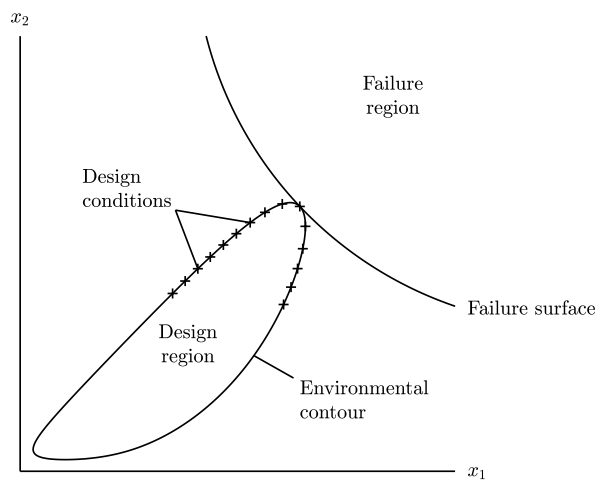

**Fig. 1.** Illustration of terminology.

Similarly, there are several reasons why it is useful to quantify the marginal exceedance probabilities associated with ISORM and HD contours. Firstly, quantifying the marginal exceedance probabilities for the highest values of each variable along an ISORM or HD contour provides a direct comparison to IFORM and DS contours. In particular, the IFORM contour is completely contained within the ISORM contour, so the return period of the largest value of each variable along the contour is a useful metric for quantifying the difference between the methods.

Secondly, the largest values of each variable along a contour provide a direct comparison to univariate analyses. Univariate analyses of extreme environmental conditions are often conducted in parallel with multivariate analyses. Moreover, some design standards for marine structures specify reliability criteria in terms of both univariate return values and environmental contours, depending on the load case (e.g. Ref. [14,15]). It is therefore of interest to examine how the maximum value of each variable along the contour compares to the univariate return value.

Comparing the marginal and total exceedance probabilities of different contour methods provides an indication of the conservatism of each method. However, the actual conservatism or non-conservatism in failure probabilities estimated using each method is dependent on the structure of interest and the environmental conditions. In this article we provide some simple examples to illustrate the impacts that the choice of contour has on estimates of extreme structural response.

The estimation of the joint distribution of environmental variables is the focus of ongoing research. Methods for modelling the joint distribution include hierarchical conditional models [4,16–19], copula models [20–23], kernel density estimates [24,25] and conditional extreme value models [26–29]. In the following we assume that a model has been estimated for the joint distribution and contours are to be derived from the joint distribution. There is a wide range of definitions of environmental contours, such as the method proposed by Haver [16,30], and the isodensity contour methods described by NORSOK [[31], p.12] and DNV GL [[32], Section 3.7.2.4], methods based on joint exceedance regions [27], or methods based on univariate analyses [33]. In the present work we focus on four types of contour, namely IFORM, ISORM, DS and HD contours, due to their analogous definitions in terms of either marginal or total exceedance probabilities.

The article is organised as follows. We start by presenting the definitions of the various types of contour in Section 2. The marginal exceedance probabilities associated with the largest values of each variable on IFORM, ISORM and HD contours are discussed in Section 3. The complementary problem of quantifying the total probability outside IFORM and DS contours is discussed in Section 4. Section 5 presents three case studies, to illustrate the effect of using different types of contour in some simplified structural design problems. Finally, a summary and conclusions are presented in Section 6.

## 2. Environmental contour definitions

Before presenting the definitions of the environmental contour methods, it is useful to introduce some notation. Consider the joint distribution of a vector of $n$ random variables, $\mathbf{X} = (X_1, \ldots, X_n)$, with joint distribution $F(\mathbf{x}) = \Pr(X_1 \leq x_1, \ldots, X_n \leq x_n)$ and marginal distributions $F_j(x_j) = \Pr(X_j \leq x_j), j = 1, \ldots n,$. The corresponding marginal exceedance probabilities are denoted $Q_j(x) = 1 - F_j(x)$. In general, we will use $\alpha$ to denote an exceedance probability. The marginal exceedance probability associated with the largest value of $X_j$ along a contour is denoted $\alpha_{m,j}$ and the total probability outside the contour is denoted $\alpha_t$ (see Fig. 2). The quantile of $F_j$ at marginal exceedance probability $\alpha$, is denoted $x_{j,\alpha}$ and is the solution of $Q_j(x_{j,\alpha}) = \alpha$. The lower and upper bounds of $X_j$ along a contour at exceedance probability α are denoted $c_{j,\alpha}^L$ and $c_{j,\alpha}^U$. In many of the examples below we are interested in the upper bound only and will drop the superscript $U$. Similarly, when we are considering the values of $x_{j,\alpha}$ and $c_{j,\alpha}$ for the same variable we will drop the subscript $j$.

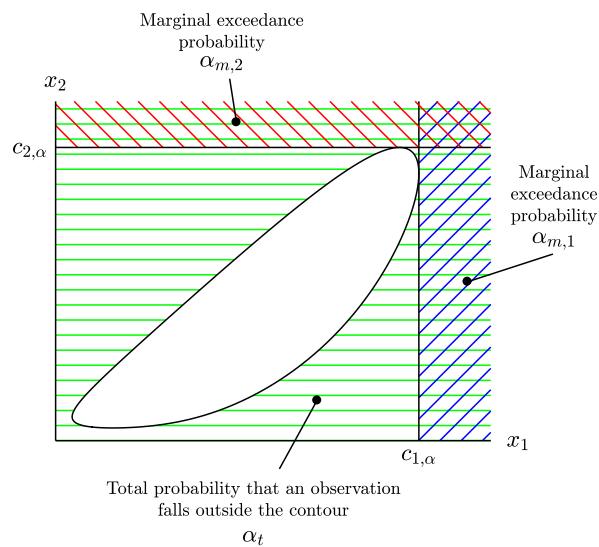

**Fig. 2.** An environmental contour with associated marginal and total exceedance probabilities.

## 2.1. Inverse first-order reliability method (IFORM)

The most commonly applied EC method in marine design is the Inverse First-Order Reliability Method (IFORM) [4]. The method is based on the First-Order Reliability Method (FORM), used to estimate a structure's failure probability [[34], pp. 71–125]. In the FORM approach, the joint distribution of $X_1, \ldots, X_n$ is transformed to independent standard normal variables, $U_1, \ldots U_n$ using the Rosenblatt transformation [35], given by

$$\begin{aligned} u_1 &= \Phi^{-1}\left(F_{X_1}(x_1)\right) \\ u_2 &= \Phi^{-1}\left(F_{X_2|X_1}(x_2|x_1)\right) \\ &\vdots \\ u_n &= \Phi^{-1}\left(F_{X_n|X_1,\ldots,X_{n-1}}(x_n|x_1,\ldots,x_{n-1})\right). \end{aligned} \quad (1)$$

where $\Phi$ is the cumulative distribution function (CDF) of the standard univariate normal distribution. The design point is defined as the highest probability point on the failure surface in $U$-space, which corresponds to the closest point of the failure surface to the origin (see Fig. 3). The radius from the origin to the design point, referred to as the reliability index, is denoted $\beta_F$. Finally, the reliability index is used to estimate the structure's probability of failure by assuming that the failure surface is linear at the design point, such that

$$\Phi(\beta_F) = 1 - \alpha. \quad (2)$$

In the IFORM approach, the location of the failure surface and design point are not known. Instead, the desired reliability index, $\beta_F$, is specified and the IFORM contour in $U$-space is defined as the set of points at radius $\beta_F$. The environmental contour in the original space is then obtained by applying the inverse Rosenblatt transformation to the contour in $U$-space.

An illustration of the definition of the IFORM contour in 1D and 2D is shown in Fig. 4 (a) and (b). Although contours are defined for two or more dimensions, the definition is still valid in 1D and serves to illustrate how the maximum value along the contour is related to the marginal return value. It is clear that for an IFORM contour at exceedance probability α, the maximum values of the standard normal variables $U_1, \ldots, U_n$ have marginal exceedance probability α in $U$-space. However, the same is not true for the maximum values of $X_1, \ldots, X_n$ along the IFORM contour in the original parameter space. In fact, the Rosenblatt transformation only ensures that the maximum value of $X_1$ along the IFORM contour, $c_{1,\alpha}$, has exceedance probability α and the exceedance probabilities of $c_{j,\alpha}$ ($j > 1$) can be either greater than or less than α. The shape of the contour and maximum values of each variable along the contour are dependent on the order in which conditional distributions are specified in the Rosenblatt transformation (2.1). This is discussed further in Section 3.1.

## 2.2. Inverse second-order reliability method (ISORM)

The IFORM approximation is conservative when the failure region is convex, but is not conservative if the failure region is concave. Chai and Leira [8] proposed a second-order approximation to the failure surface which is always conservative. In the ISORM method the failure surface is assumed to enclose a circle in $U$-space, centred at the origin. The radius $\beta_{Sn}$, is defined so that the probability that an observation falls outside the circular region is α. Chai and Leira noted that since the sum of $n$ independent standard normal variables follows a Chi-squared distribution on $n$ degrees of freedom, $\chi_n^2$, the radius $\beta_{Sn}$ can be written as:

$$\chi_n^2(\beta_{Sn}^2) = 1 - \alpha. \quad (3)$$

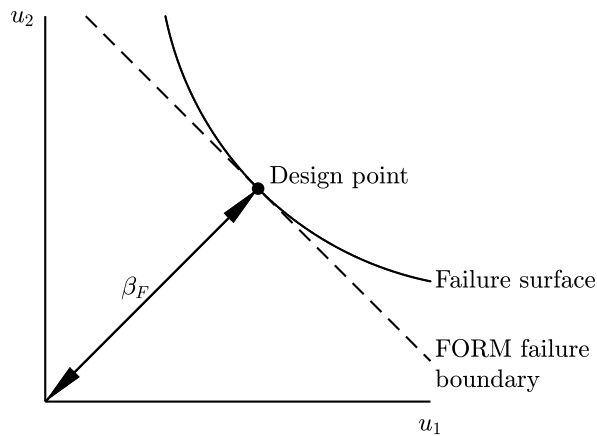

**Fig. 3.** Illustration of FORM approximation to failure surface.

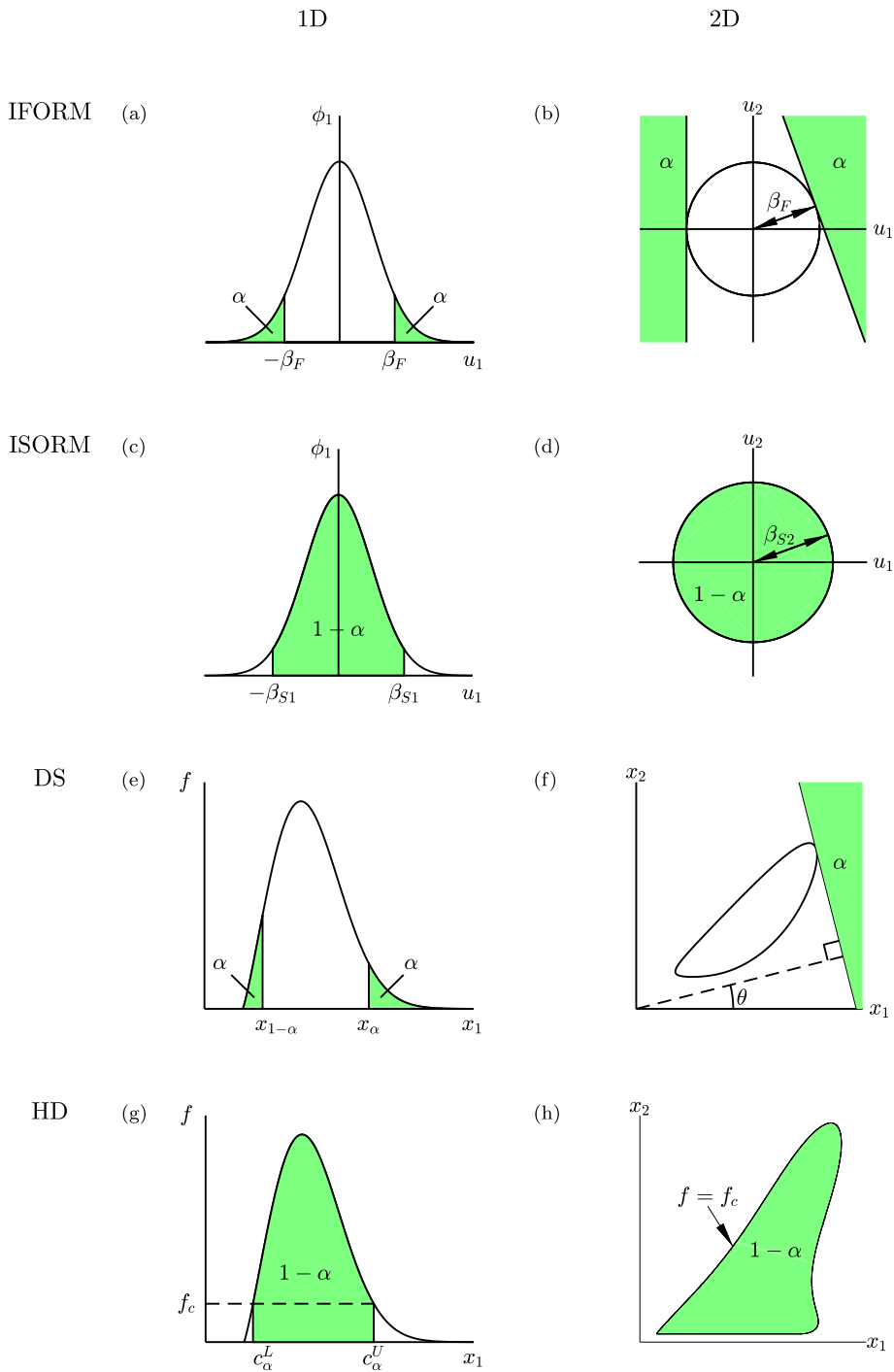

**Fig. 4.** Illustration of definitions of IFORM, ISORM, direct sampling (DS) and highest density (HD) contours in 1D and 2D.

In the ISORM definition, the radius, $\beta_{Sn}$, is a function of both the exceedance probability α and the number of dimensions, $n$. As with the IFORM contour, the ISORM contour in the original space is obtained by applying the inverse Rosenblatt transformation to the contour in $U$-space. An illustration of the definition of ISORM contours in 1D and 2D is shown in Fig. 4 (c) and (d). As with the IFORM method, ISORM contours are dependent on the order in which conditional distributions are specified in the Rosenblatt transformation. This effect is discussed further in Sections 3.1 and 3.2.

The ISORM contour definition proposed by Chai and Leira [8] represents a particular type of SORM approximation. More generally, SORM can be used to provide a less conservative approximation to a convex failure region (see e.g. Refs. [36,37]). However, in the present context we shall only consider the more conservative ISORM contours defined above.

### 2.3. Direct sampling method (DS)

An alternative to the IFORM method was proposed by Huseby et al. [7]. They noted that the assumption of a linear failure surface in *U*-space can be difficult to justify, since when the inverse Rosenblatt transformation is applied, the assumed failure surface is no longer linear in the original parameter space. In the DS method, the contour is defined in a similar way to the IFORM contour, but with the failure surface assumed to be linear in the original parameter space.

In 2D, the contours are constructed as follows. For a given angle θ, the halfspace perpendicular to the line at angle θ to the origin is found which contains probability α (see Fig. 4(f)). A finite number of angles are selected and contours are defined as the intersection of the lines bounding each halfspace. Huseby et al. proposed that the location of the halfspace at each angle could be estimated by Monte Carlo simulation under the joint distribution and projection of the variables on to the line at angle θ to the origin: $y = x_1 \cos\theta + x_2 \sin\theta$. The distance of the halfspace from the origin is then found as the empirical quantile of *y* at exceedance probability α.

The method can be interpreted as a rotation of the axes and calculation of a marginal exceedance probability on the rotated axes (this is also true of the IFORM method in *U*-space). The use of Monte Carlo simulation to estimate the marginal exceedances avoids numerical issues with integrating the joint distribution and allows the method to be easily extended into higher dimensions.

We see that, by definition, the maximum value of each variable along a DS contour has marginal exceedance probability α. Since DS contours are defined as the intersection of straight line segments which describe exceedances at a given angle, the resulting contours define a design region which is always convex, regardless of the shape of the joint distribution.

### 2.4. Highest density method (HD)

Highest density (HD) contours can be thought of as the *X*-space analogue of ISORM contours in *U*-space, in the same way that DS contours are *X*-space analogues of IFORM contours *U*-space. HD contours were proposed as a conservative environmental contour that does not assume a convex failure region [9]. They are defined based on highest density regions, a statistical concept used in a range of contexts (see e.g. Ref. [38]). A highest density region, *R*, is the smallest possible region in the variable space that contains a given probability content. Mathematically, it can be expressed as the set of all **x** whose probability density is greater than a threshold $f_c$:

$$R(f_c) = \{\mathbf{x} \in \mathbb{R}^n : f(\mathbf{x}) \geq f_c\}, \qquad (4)$$

where $f(\mathbf{x})$ is the joint density function and $f_c$ is chosen as the largest threshold that yields a region, which contains a probability of at least $1-\alpha$, that is

$$f_c = \underset{f \in [0,\infty)}{\mathrm{argmax}} \Pr(\mathbf{X} \in R(f)) \geq 1 - \alpha. \qquad (5)$$

Then, the α-exceedance highest density contour is defined as the set $C(\alpha) \subset R(f_c)$ that contains exactly the environmental states at which the probability density equals $f_c$:

$$C(\alpha) = \{\mathbf{x} \in \mathbb{R}^n : f(\mathbf{x}) = f_c\}. \qquad (6)$$

### 2.5. Rationale for definition of contours in terms of marginal and total exceedance probabilities

For the contour methods discussed above, the definitions of the exceedance regions are related to the assumptions made about the shape of the failure region. Since the shape of the failure region is not known when a contour is constructed, the conservative approximations for the shape of the failure surface result in the inclusion of both severe and non-severe conditions in the exceedance

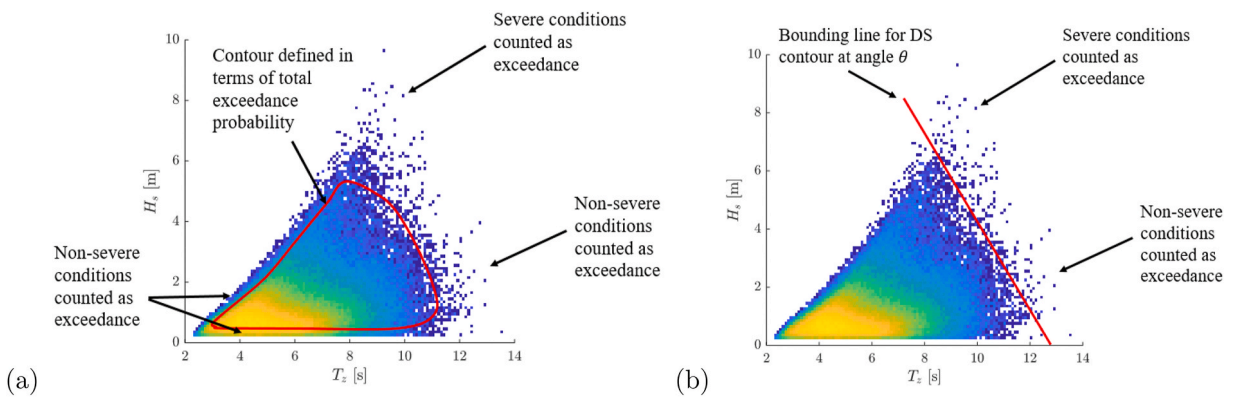

**Fig. 5.** Sketch showing which observations are counted as exceedances for (a) ISORM or HD contours, and (b) DS contours.

region. This is a feature of all the contour methods described above.

Consider the sketches in Fig. 5, which show 10 years of observations of significant wave height, $H_s$, and zero-up-crossing period, $T_z$ from NDBC wave buoy 44,007, located off the coast of Maine, USA. Fig. 5(a) shows a contour defined in terms of the total exceedance probability. For this contour, observations with $H_s$ both above the contour upper bound and below the contour lower bound are included in the set of exceedances. Fig. 5(b) shows one of the bounding lines used to construct DS contours at some high exceedance probability. The observations to the right of the line are all counted as exceedances at this particular angle. The set of exceedances for this angle includes both conditions with high $H_s$, which may contribute to the failure probability of a structure, and conditions with high $T_z$ but low $H_s$, which are less likely to contribute to the failure probability. Thus, the difference between contours defined in terms of marginal and total exceedance probabilities is not whether they include non-severe conditions in the set of exceedances, but which non-severe conditions are defined to be exceedances.

In the environmental contour method the location and shape of the structure's failure surface is not known (in fact the structure of interest need not even be specified in order to construct contours). The inclusion of non-severe conditions in the set of exceedances results from either the conservative linear approximation to a convex failure region in the IFORM and DS methods, or the conservative approximation of a concave failure region enclosing the design region in the ISORM and HD methods.

If the HD and ISORM methods are applied to calculate contours for $H_s$ and $T_z$, then the rationale is not to assume that the structure fails at low values of $H_s$. Rather, the rationale is to define some region of the parameter space, which encloses a total probability of $1 - \alpha$, so that no assumption about the shape of the failure region is required. Similarly, in the example shown in Fig. 5(b), the rationale for defining exceedances in this way is to provide a linear approximation to a failure surface, without the need to specify where the failure surface is located.

It would be straightforward to extend the definition of HD or DS contours to define a region of parameter space containing a specified probability which does not exclude low values of either $X_1$ or $X_2$, or both (it is less straightforward to do this for IFORM and ISORM contours due to the use of the Rosenblatt transformation). However, this would require setting a threshold for what constitutes a "low" value, which would be somewhat arbitrary. There are infinitely many ways to define a region which contains probability $1 - \alpha$. In this work we focus on the four contour methods described above since their definitions are quite general and do not require any subjective judgements about threshold levels.

## 3. Marginal exceedance probabilities

### 3.1. IFORM contours

As noted in Section 2.1, the shape of the IFORM contour and the marginal exceedance probabilities of $c_{j,\alpha}$ depend on the order in which the Rosenblatt transformation is applied. The effect of this is illustrated through two examples below.

#### 3.1.1. Example 1

Define a joint probability density function (PDF) as the sum of two bivariate normal distributions, with mean values centred at $\pm (a, a)$:

$$f_{X_1,X_2}(x_1, x_2) = \frac{1}{2}[\varphi(x_1 - a)\varphi(x_2 - a) + \varphi(x_1 + a)\varphi(x_2 + a)], \tag{7}$$

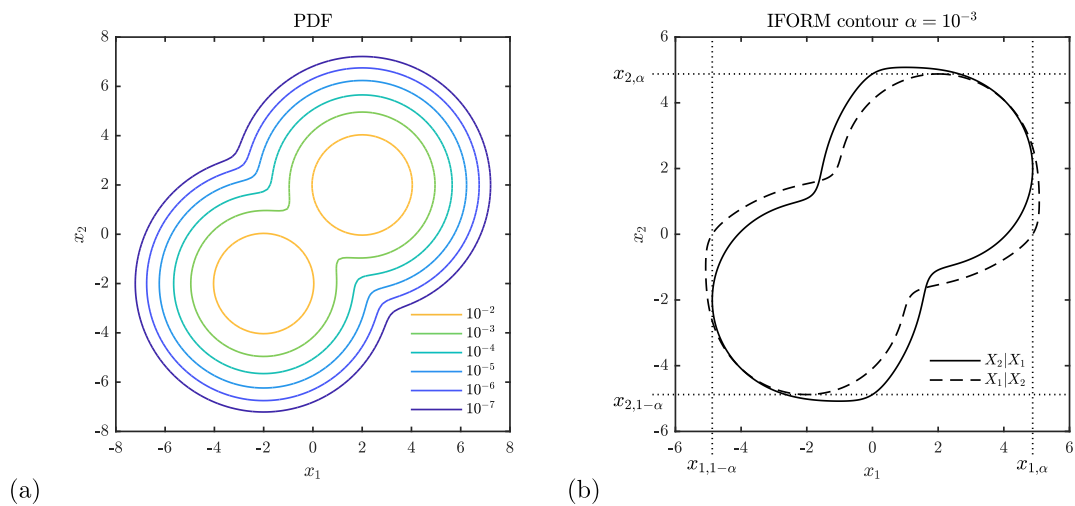

**Fig. 6.** (a) Isodensity contours for joint PDF given in (7). (b) IFORM contours at exceedance probability $\alpha = 10^{-3}$ for joint distribution given in (7), based on Rosenblatt transformation specified in terms of $X_2|X_1$ or $X_1|X_2$. Marginal return values at exceedance $\alpha$ and $1 - \alpha$ for each variable shown as dotted lines.

where $\varphi$ is the PDF of the standard normal distribution. The joint PDF $f_{X_1,X_2}(x_1, x_2)$ is illustrated in Fig. 6(a) for the case $a = 3$. The PDF is symmetric about the line $x_1 = x_2$. The marginal CDF of $X_1$ is

$$F_{X_1}(x_1) = \frac{1}{2}[\Phi(x_1 - a) + \Phi(x_1 + a)], \tag{8}$$

and the conditional of CDF of $X_2|X_1$ is given by

$$F_{X_2|X_1}(x_2|x_1) = \frac{\varphi(x_1 - a)\Phi(x_2 - a) + \varphi(x_1 + a)\Phi(x_2 + a)}{\varphi(x_1 - a) + \varphi(x_1 + a)}. \tag{9}$$

Through the symmetry of (7), the marginal CDF of $X_2$ and conditional CDF of $X_1|X_2$ are obtained by switching the variables in the expressions above. IFORM contours can be derived by applying the Rosenblatt transformation as defined in (2.1) or by changing the conditioning variable to be $X_2$ and defining $u_1 = \Phi^{-1}(F_{X_2}(x_2))$, $u_2 = \Phi^{-1}(F_{X_1|X_2}(x_1|x_2))$. The resulting IFORM contours for the two transformations are shown in Fig. 6(b) for an exceedance probability of $\alpha = 10^{-3}$. Despite the symmetry of the PDF in the line $x_1 = x_2$, the IFORM contours are not symmetric about this line and the minimum and maximum values of each variable along the contours differ. The contour defined using the transformation in terms of $X_2|X_1$ passes through the points $(x_{1,\alpha}, \text{median}(x_2|x_{1,\alpha}))$ and $(x_{1,1-\alpha}, \text{median}(x_2|x_{1,1-\alpha}))$. However, the minimum value of $X_2$ on the contour is less than $x_{2,1-\alpha}$ and the maximum value of $X_2$ is greater than $x_{2,\alpha}$. In general, if $X_1$ is used as the conditioning variable in the Rosenblatt transformation, then there is no fixed relationship between the maximum and minimum values of $X_2$ along the contour and the marginal return values of $X_2$. This is illustrated in the next example.

### 3.1.2. Example 2

In this example, suppose that $X_1$ follows a Weibull distribution, with CDF

$$F_{X_1}(x_1) = 1 - \exp\left(-\left(\frac{x_1 - \gamma}{\lambda}\right)^k\right). \tag{10}$$

Suppose also that the distribution of $X_2$ conditional on $X_1$ is normal, with mean $\mu = a_1 + a_2 x_1$ and standard deviation $\sigma = b_1 + \exp(b_2 x_1)$. The joint PDF can be written

$$f_{X_1,X_2}(x_1, x_2) = f_{X_1}(x_1) f_{X_2|X_1}(x_2|x_1). \tag{11}$$

To express the joint PDF as $f_{X_1,X_2}(x_1, x_2) = f_{X_2}(x_2) f_{X_1|X_2}(x_1|x_2)$ requires numerical integration and differentiation. However, it is straightforward to apply the Rosenblatt transformation numerically using either $X_1$ or $X_2$ as the conditioning variable.

An example is shown in Fig. 7 for the case $\gamma = 0$, $\lambda = 1$, $k = 3$, $a_1 = 0$, $a_2 = 1$, $b_1 = 0.1$, $b_2 = -1$. It is evident that the shape of the contour is dependent on the order in which the Rosenblatt transformation is applied. Moreover, the minimum and maximum values of the conditional variable along the contour are not in fixed relation to the marginal return values. For the contours based on the transformation in terms of $X_2|X_1$ (solid lines), the contour at $\alpha = 10^{-1}$ has $c_{2,\alpha}^L < x_{2,1-\alpha}$ and $c_{2,\alpha}^U < x_{2,\alpha}$, whereas for the contour at $\alpha = 10^{-5}$ we have $c_{2,\alpha}^L < x_{2,1-\alpha}$ and $c_{2,\alpha}^U > x_{2,\alpha}$.

In this example, the differences between the IFORM contours using the different transformation orders are relatively small. Nevertheless, it is important to be aware that for a given joint PDF, IFORM contours are not uniquely defined. The impact that this has on estimates of structural responses is discussed in Section 5.

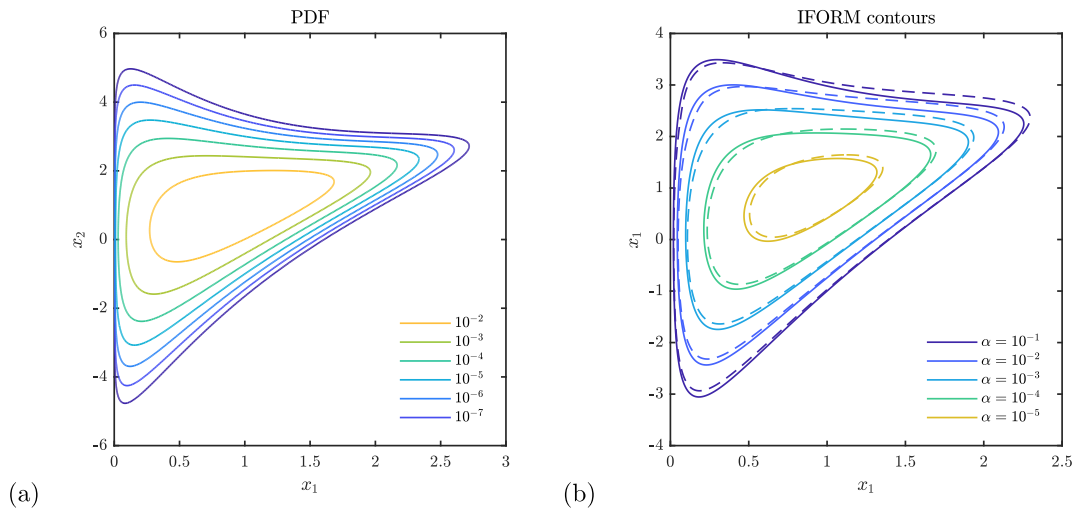

**Fig. 7.** (a) Isodensity contours of joint PDF in Example 2. (b) IFORM contours for Example 2, based on Rosenblatt transformation specified in terms of $X_2|X_1$ (solid lines) or $X_1|X_2$ (dashed lines).

## 3.2. ISORM contours

Since IFORM and ISORM contours only differ in the value of β used to define the contour in $U$-space, the discussion of the influence of the transformation on IFORM contours in Section 3.1 is also applicable to ISORM contours. That is, for an ISORM contour at exceedance probability $\alpha_t$, only $\alpha_{m,1} = Q(c_{1,\alpha_t})$ is in fixed relation to $\alpha_t$ and $\alpha_{m,j}$ ($j > 1$) depends on the shape of the joint distribution. In the following, we shall only consider the relation between $\alpha_{m,1}$ and $\alpha_t$.

The ratio $\alpha_t/\alpha_{m,1}$ for an ISORM contour in $n$ dimensions can be calculated as follows. For a given contour exceedance probability, $\alpha_t$, the radius, $\beta_{Sn}$, of the ISORM contour in $U$-space is given by (3). The marginal exceedance probability, of $c_{1,\alpha_t}$ is then given by

$$\alpha_{m,1} = 1 - \Phi(\beta_{Sn}) = 1 - \Phi\left(\sqrt{[\chi_n^2]^{-1}(1-\alpha_t)}\right), \tag{12}$$

where $[\chi_n^2]^{-1}$ is the inverse of the chi-squared CDF. The ratio $\alpha_t/\alpha_{m,1}$ for ISORM contours is therefore a function of $\alpha_t$ and $n$. At present environmental contours are most commonly used in two or three dimensions. However, in principle, the method is applicable to any number of dimensions. Fig. 8 shows the ratio $\alpha_t/\alpha_{m,1}$ against $\alpha_t$ for $n = 1,...,5$. For a one dimensional ISORM contour, the exceedance probability is split evenly between the upper and lower tails of the distribution, so the contour has an non-exceedance probability of $\alpha_t/2$ in the lower tail and an exceedance probability of $\alpha_t/2$ in the upper tail. Consequently, in 1D we have $\alpha_t/\alpha_{m,1} = 2$. For higher dimensions, the ratio $\alpha_t/\alpha_{m,1}$ increases with both $\alpha_t$ and the number of dimensions. For an ISORM contour at exceedance probability $\alpha_t = 10^{-3}$ we have $\alpha_t/\alpha_{m,1} \approx 10$ in two dimensions, and $\alpha_t/\alpha_{m,1} \approx 115$ in four dimensions.

The ratio $\alpha_t/\alpha_{m,1}$ is the ratio of the return period of $c_{1,\alpha_t}$ to the contour return period. For example, if sea states are reported at 3 h intervals (assuming stationary conditions over this interval), the 50-year contour corresponds to an exceedance probability of $\alpha_t = 1/(50 \times 365.25 \times 24/3) = 6.84 \times 10^{-6}$. So the 50-year 2D ISORM contour passes through points with marginal return periods of 635 years. Similarly, the 50-year 4D ISORM contour passes through points with marginal return periods of 10,950 years. Or to put it another way, since IFORM contours are defined in terms of marginal probabilities, we can say that for 3-h observations of 2 variables the 50-year ISORM is equivalent to the 635-year IFORM contour and for 4 variables the 50-year ISORM contour is equivalent to the 10,950-year IFORM contour.

The difference between $c_{1,\alpha_t}$ and $x_{1,\alpha_t}$ depends on the shape of the tail of the distribution. For long-tailed distributions a small change in return period results in a large change in the return value, whereas for short-tailed distributions a large change in return-period will only result in a small change in return value. An example is given here assuming that $X_1$ follows a Weibull distribution (10). Fig. 9 shows the Weibull distribution for cases with location parameter $\gamma = 0$, scale parameter $\lambda = 1$ and various shape parameters $k$. When $k = 1$ the distribution has an exponential tail. As $k$ increases the upper tail becomes shorter. The ratio between the largest value of $X_1$ on the 50-year ISORM contour and the 50-year marginal return value (denoted $c_{50}$ and $x_{50}$, for brevity) is shown in Fig. 10 as a function of the Weibull shape parameter, $k$ for contours in $n = 1,...,5$ dimensions. Here we assume a location parameter $\gamma = 0$ (the ratio $c_{50}/x_{50}$ is invariant to scale, $\lambda$) and assume observations at 3-h intervals as above. The ratio $c_{50}/x_{50}$ decreases as the Weibull shape parameter increases and the upper tail becomes shorter. For a 2D contour $c_{50}$ is 21% larger than $x_{50}$ when $k = 1$ and 10% larger when $k = 2$. For a 4D contour $c_{50}$ is 45% larger than $x_{50}$ when $k = 1$ and 20% larger when $k = 2$. These differences can potentially have a large impact on the strength requirements for a structure, so the decision between the use of an ISORM contour over an IFORM or DS contour needs to be carefully justified. Some examples where each type may appropriate are presented in Section 5.

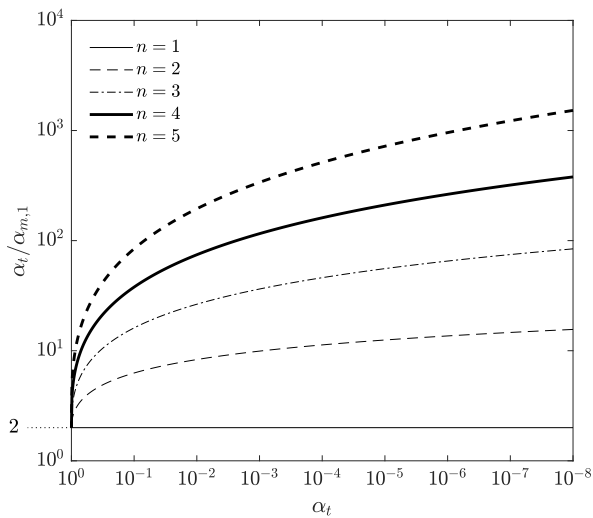

**Fig. 8.** Ratio of total exceedance probability, $\alpha_t$, to marginal exceedance probability of largest value of $X_1$ along the contour, $\alpha_{m,1} = Q(c_{1,\alpha_t})$, against $\alpha_t$ for ISORM contours in $n = 1, ..., 5$ dimensions.

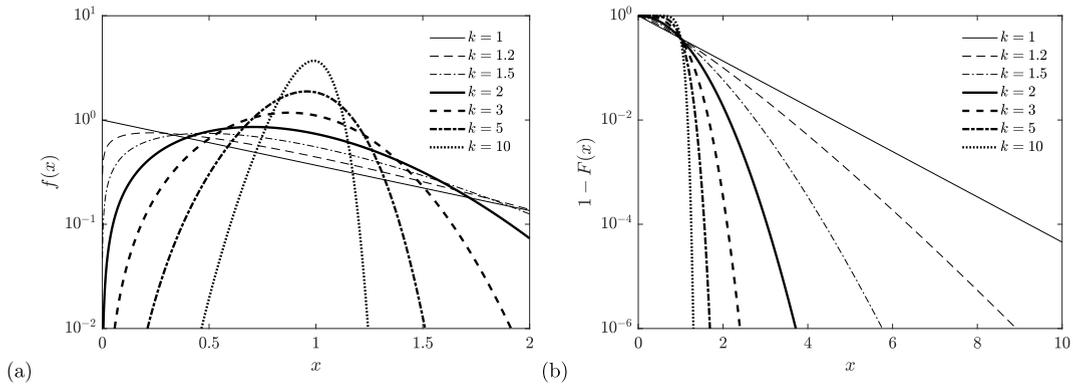

**Fig. 9.** (a) PDF of the Weibull distribution. (b) Upper tail of the Weibull distribution function. Both plots for location parameter $\gamma = 0$, scale parameter $\lambda = 1$ and various shape parameters $k$.

### 3.3. HD contours

HD contours are based on the same notion of exceedance as ISORM contours (i.e. they are defined in terms of the total probability outside the contour, $\alpha_t$), so we may expect a similar relationship between the total exceedance probability and the marginal exceedance probabilities of $c_{j,\alpha_t}$. For HD contours in any number of dimensions we have, by definition, $c_{j,\alpha_t} \geq x_{j,\alpha_t}$, since it is not possible for a region of parameter space bounded by $x_{j,\alpha_t}$ in one dimension to contain a probability greater than $\alpha_t$.

However, as the HD contour is defined as an isodensity contour in the original parameter space, the relationship between $Q(c_{j,\alpha_t})$ and $\alpha_t$ is dependent on the shape of the density function. Consider the 1D case shown in Fig. 4 (g). In the example shown it is clear that $F(c_\alpha^L) < Q(c_\alpha^U)$. In general we have $F(c_\alpha^L) < Q(c_\alpha^U)$ when the lower tail is steeper than the upper tail, $F(c_\alpha^L) > Q(c_\alpha^U)$ when the upper tail is steeper than the lower tail and $F(c_\alpha^L) = Q(c_\alpha^U)$ when the distribution is symmetric. So the non-exceedance probability of the contour lower bound and exceedance probability of the contour upper bound are not in fixed relation to the contour exceedance probability.

Consider the hypothetical cases where the PDF has a triangular shape, shown in Fig. 11. In the extreme case, where the lower tail has a vertical gradient, the α-exceedance probability HD contour has zero non-exceedance probability in the lower tail and exceedance probability α in the upper tail, so that $c_\alpha = x_\alpha$ (i.e. the upper bound of the α-exceedance contour is equal to the α-exceedance marginal quantile). In the case that the PDF is symmetric, the exceedance probability is split evenly between the lower and upper tails so that $c_\alpha = x_{\alpha/2}$ (i.e. the upper bound of the α-exceedance contour is equal to the marginal quantile at exceedance probability $\alpha/2$). In the other extreme case, where the upper tail has a vertical gradient and an upper end point at $x_m$, the exceedance probability of the upper bound of the HD contour is zero and $c_\alpha = x_m$ (i.e. the upper bound of the contour is the upper end point of the distribution).

These examples are unrealistic, but serve to illustrate the sensitivity of the minimum and maximum points on the contour to the shape of the density function. Now consider a more realistic 1D example, where the random variable $X_1$ follows a Weibull distribution. The PDF of the Weibull distribution is

$$f_{X_1}(x_1) = \frac{k}{\lambda}\left(\frac{x_1-\gamma}{\lambda}\right)^{k-1}\exp\left[-\left(\frac{x_1-\gamma}{\lambda}\right)^k\right]. \tag{13}$$

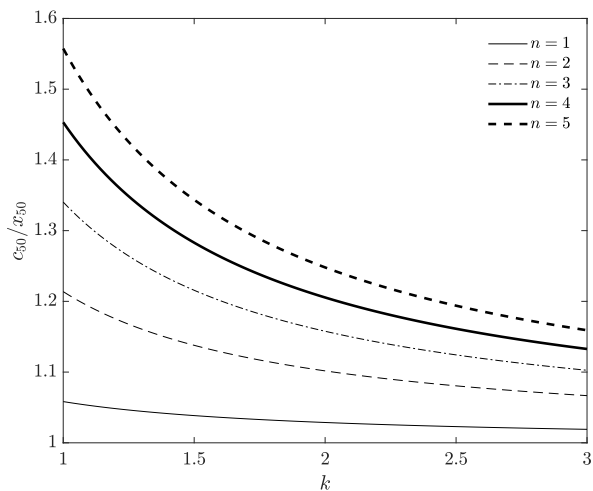

**Fig. 10.** Ratio of the largest value of $X_1$ on a 50-year ISORM contour, $c_{50}$, to the 50-year marginal return value, $x_{50}$, assuming a marginal Weibull distribution with location parameter $\gamma = 0$ and shape parameter $k$. Results shown for ISORM contours in $n = 1, \ldots, 5$ dimensions.

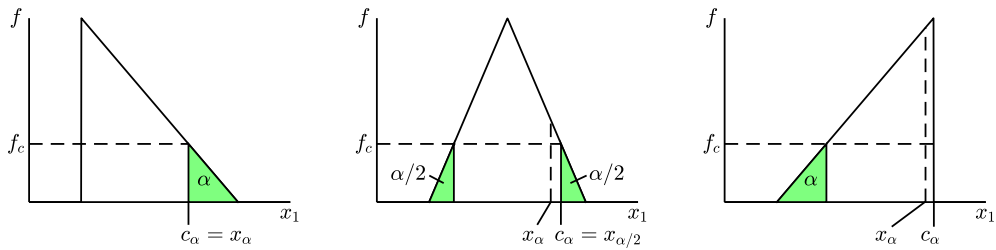

**Fig. 11.** Examples of marginal and contour exceedance levels for HD method applied to 1D triangular PDFs.

When $k=1$ the PDF has a vertical gradient in the lower tail and a long upper tail (see Fig. 9 (a)). As $k$ increases, the gradient of the lower tail decreases and the gradient of the upper tail increases. As $k \to \infty$ the PDF converges towards a Dirac delta distribution centred at $X_1 = \lambda + \gamma$. However, for large finite $k$ the gradient of the lower tail is less than the gradient of the upper tail.

The effect that the shape of the PDF has on the upper bound of the contour is assessed as follows. For a given total exceedance probability $\alpha_t$, the $\alpha_t$ - exceedance HD contour is calculated and the upper bound of the contour $c_{\alpha_t}$ is found. The ratio $c_{\alpha_t}/x_{\alpha_t}$ of the contour upper bound to the marginal quantile at exceedance probability $\alpha_t$ is shown in Fig. 12(a) as a function of Weibull shape parameter $k$ for various contour exceedance probabilities $\alpha_t$. When $k=1$ we have $c_{\alpha_t} = x_{\alpha_t}$ at all exceedance probabilities, since the lower tail of the PDF has a vertical gradient. As $k$ increases the ratio $c_{\alpha_t}/x_{\alpha_t}$ increases as the gradients in the upper and lower tail change. The ratio $c_{\alpha_t}/x_{\alpha_t}$ is dependent on the exceedance probability, since the gradients in the tails are not constant, with $c_{\alpha_t} \to x_{\alpha_t}$ as $\alpha_t \to 0$. As $k$ increases and the upper tail becomes shorter than the lower tail, $c_{\alpha_t} \to x_{\alpha_t}$ since the value of $x_{\alpha_t}$ becomes less sensitive to the exceedance probability $\alpha_t$.

The ratio of the contour exceedance probability to the exceedance probability of the largest point on the contour, $\alpha_t/Q(c_{\alpha_t})$, is shown in Fig. 12(b). In this case the ratio increases monotonically with $k$. For smaller values of $k$ the ratio $\alpha_t/Q(c_{\alpha_t})$ is larger for larger values of $\alpha_t$, but at higher values of $\alpha_t$ the trend is reversed. Although $\alpha_t/Q(c_{\alpha_t})$ can be large for large $k$, the difference between $c_{\alpha_t}$ and $x_{\alpha_t}$ is small since the distribution has a steep gradient in the upper tail.

In these examples we see that the relationship between the largest value of each marginal variable on the HD contour and the marginal quantile at the same exceedance probability is relatively insensitive to the shape of the distribution, despite large differences in the exceedance probability.

In two dimensions the relationship between the marginal exceedance probability of the maximum point along a contour and contour exceedance probability is slightly different. Consider the following simple cases where $X_1$ follows a Weibull distribution with $\gamma=0$, $\lambda=1$ and $k=1$ and $X_2$ is independent of $X_1$, and is either uniformly distributed over [-3 3] or follows a standard normal distribution. The joint PDF for the two cases are illustrated in Fig. 13, together with the HD contour at $\alpha_t = 0.1$ and the marginal quantile $x_{1,\alpha_t}$. As the Weibull shape parameter is 1, the joint PDF has a vertical gradient at $x_1 = 0$. In the case where $X_2$ is uniformly distributed the $\alpha_t$-exceedance HD contour is simply a line at the marginal quantile $x_{1,\alpha_t}$ and the 2D contour is equivalent to the 1D case. However, when the joint density contours are no longer straight lines parallel to the $X_2$ axis, the highest value of $x_1$ along the $\alpha_t$-exceedance HD contour will be greater than $x_{1,\alpha_t}$, since the region $X_1 \le x_{1,\alpha_t}$ is no longer the highest density region containing a probability $1-\alpha_t$. So in two dimensions, the criteria that one variable has a vertical gradient in the lower tail of the PDF is no longer sufficient to ensure that $c_{j,\alpha_t} = x_{j,\alpha_t}$ and in general we will have $c_{j,\alpha_t} > x_{j,\alpha_t}$.

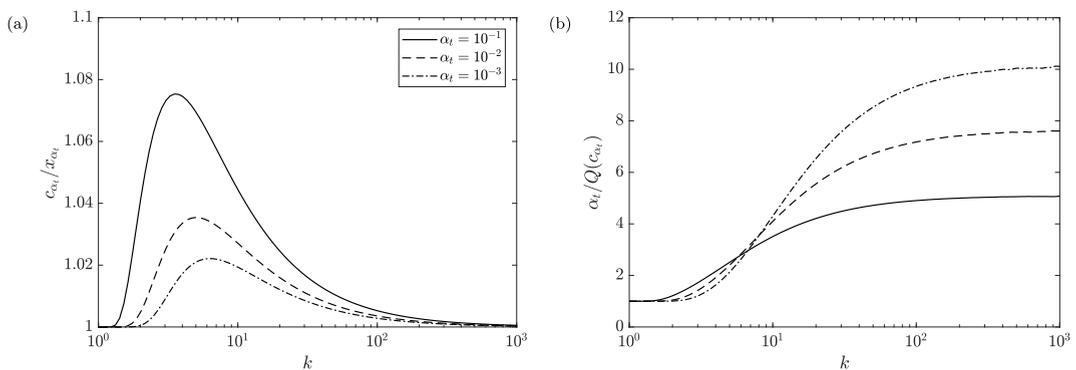

**Fig. 12.** (a) Ratio $c_{\alpha_t}/x_{\alpha_t}$ for a 1D HD contour. (b) Ratio of contour exceedance probability to marginal exceedance probability of largest value of $X_1$ along contour. Both ratios shown as a function of Weibull shape parameter, $k$ for various total exceedance probabilities $\alpha_t$.

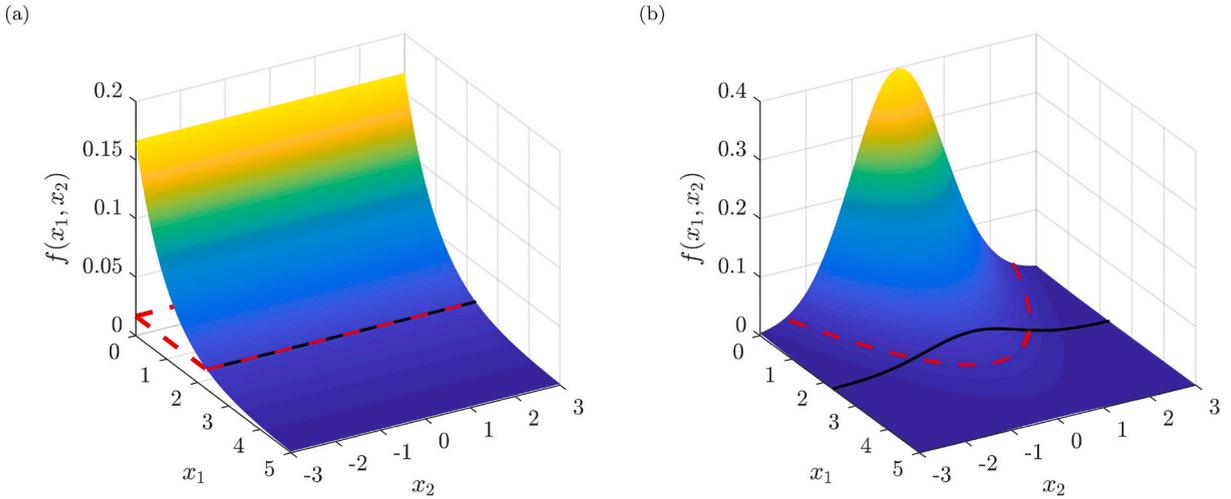

**Fig. 13.** (a) Joint PDF where $X_1$ is Weibull distributed with shape parameter $k = 1$ and $X_2$ is independent and uniformly distributed. (b) Joint PDF where $X_1$ is Weibull distributed with shape parameter $k = 1$ and $X_2$ is independent and normally distributed. Black lines indicate marginal quantile of $X_1$ at $\alpha = 0.1$. Red dashed lines indicates HD contour at $\alpha = 0.1$. In case (a) the HD contour coincides with the marginal quantile. (For interpretation of the references to colour in this figure legend, the reader is referred to the Web version of this article.)

The effect of the Weibull shape parameter on the ratios $c_{1,\alpha_t}/x_{1,\alpha_t}$ and $\alpha_t/Q(c_{1,\alpha_t})$ is illustrated in Fig. 14 for the case where $X_1$ is Weibull distributed and $X_2$ is independent of $X_1$ and follows a standard normal distribution. In contrast to the 1D case, the ratio $c_{1,\alpha_t}/x_{1,\alpha_t}$ for the 2D contour is monotonically decreasing with both the shape parameter $k$ and contour exceedance probability $\alpha_t$. As noted above, when $k = 1$, we have $c_{1,\alpha_t} > x_{1,\alpha_t}$. As $k$ increases and the gradient of the upper tail increases and the gradient of the lower tail decreases, the position of the HD contour is shifted towards higher values of $X_1$ and the upper bound of the contour tends toward $x_{1,\alpha_t}$. In general, the differences between the 1D case and 2D case will be dependent on the shape of the joint distribution. In the case that $X_2$ is uniformly distributed, the 2D case is identical to the 1D case (in terms of the upper and lower bounds of the contour). As the distribution of $X_2$ moves away from being uniform, the differences between the 1D case and 2D case will increase.

To illustrate the relationship between marginal and total exceedance probabilities and quantiles in a more realistic case, we use a sea state model that has been used in several previous studies on environmental contours (e.g. Refs. [7–9,39]). The model describes the bivariate distribution of $H_s$ and $T_z$. The joint density function is written as

$$f_{H_s,T_z}(h_s, t_z) = f_{H_s}(h_s) f_{T_z|H_s}(t_z|h_s), \quad (14)$$

where $H_s$ is assumed to follow a translated Weibull distribution (13) and $T_z$ follows a log-normal distribution conditional on $H_s$:

$$f_{T_z|H_s}(t_z|h_s) = \frac{1}{t_z \sigma \sqrt{2\pi}} \exp\left[\frac{-(\ln t_z - \mu)^2}{2\sigma^2}\right] \quad (15)$$

The Weibull distribution's parameters are $\lambda = 2.776$, $k = 1.471$ and $\gamma = 0.8888$ while the conditonality of $T_z|H_s$ is modelled with

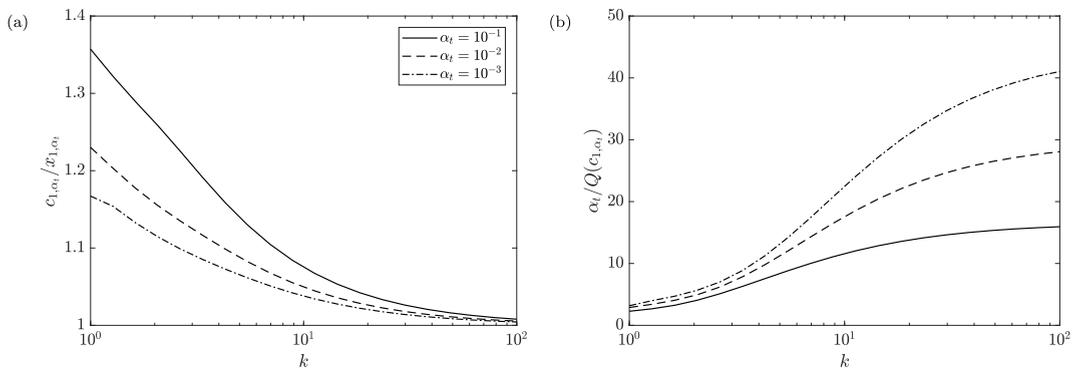

**Fig. 14.** (a) Ratio $c_{1,\alpha_t}/x_{1,\alpha_t}$ for a 2D HD contour, where $X_1$ is Weibull distributed and $X_2$ is independent and normally distributed. (b) Ratio of contour exceedance probability to marginal exceedance probability of largest value of $X_1$ along the contour. Both ratios shown as a function of Weibull shape parameter, $k$ for various total exceedance probabilities $\alpha_t$.

two dependence functions:

$$\mu = 0.1 + 1.489 h_s^{0.1901},$$
$$\sigma = 0.04 + 0.1748\exp(-0.2243 h_s). \qquad (16)$$

Fig. 15(a) shows the computed HD contours at exceedance probabilities $\alpha_t \in [10^{-7}, 10^{-1}]$. The influence of the Weibull shape parameter on the 2D contours was also investigated. Fig. 15(b) shows the HD contours at an exceedance probability of $\alpha_t = 10^{-5}$ for distributions with various Weibull shape parameters (all other parameters of the joint distribution have been left unchanged).

For comparison with the 1D case, we will consider the maximum value of $H_s$ along the HD contour, as the marginal distribution of $H_s$ is Weibull. Fig. 16(a) shows the ratio $c_{\alpha_t}/x_{\alpha_t}$, of the maximum value of $H_s$ along the contour to the marginal value of $H_s$ at the same exceedance probability. The ratio is shown as a function of the contour exceedance probability $\alpha_t$ for values of $k$ between 1 and 3. As in the 1D case, the ratio is larger at higher exceedance probabilities. The variation of $c_{\alpha_t}/x_{\alpha_t}$ is similar to the simple case above, where $X_2$ is independent and normally distributed, with the ratio decreasing as $k$ increases and the marginal distribution of $X_1$ has a steeper gradient in the upper tail. At an exceedance probability of $\alpha_t = 10^{-6}$ the HD contour maximum $H_s$ is approximately 5% larger than the marginal quantile when $k = 3$ and approximately 10% larger when $k = 1$. However, in the 1D case we had $c_{\alpha_t} = x_{\alpha_t}$ when $k = 1$, whereas in 2D $c_{\alpha_t} > x_{\alpha_t}$ when $k = 1$, despite the gradient of lower tail of the distribution being vertical at the lower end point of the distribution.

Fig. 16(b) shows the ratio $\alpha_t/Q(c_{\alpha_t})$, of the total exceedance probability to marginal exceedance probability of highest $H_s$ value along contour. As in the 1D case, the ratio of probabilities increases as the exceedance probability decreases and also as the tail becomes shorter. For a total exceedance probability of $10^{-6}$ the marginal exceedance probability of the highest $H_s$ along the contour is approximately 4 times lower when $k = 1$ and approximately 13 times lower when $k = 3$.

## 4. Total exceedance probabilities

### 4.1. IFORM contours

We now consider the complementary problem of calculating the probability that an observation falls anywhere outside a contour. For an IFORM contour defined at marginal exceedance probability $\alpha_m$, the radius of the contour in $U$-space, $\beta_F$, is given by (2). The total probability outside the contour, $\alpha_t$, is then given by

$$\alpha_t = 1 - \chi_n^2(\beta_F^2) = 1 - \chi_n^2\left(\left[\Phi^{-1}(1-\alpha_m)\right]^2\right). \qquad (17)$$

The relationship between $\alpha_m$ and $\alpha_t$ derived here is the same as in Section 3.2. However, for ease of interpretation, the ratio $\alpha_t/\alpha_m$ is plotted against $\alpha_m$ in Fig. 17 for $n = 1,...,5$. Note that the lowest possible marginal exceedance probability for an IFORM contour is 0.5 and in this case the corresponding total exceedance probability is 1. Therefore all the lines in Fig. 17 converge at $\alpha_m = 0.5, \alpha_t/\alpha_m = 2$. In two dimensions the ratio $\alpha_t/\alpha_m = O(10)$ for $\alpha_m \leq 10^{-3}$. This means that the probability of observing an exceedance anywhere outside the contour is approximately 10 times higher than the probability of observing an exceedance in any particular half-plane tangential to the contour in $U$-space (i.e. a marginal exceedance at probability $\alpha_m$ under some rotation of the axes). For a given marginal exceedance probability, the probability of observing an exceedance anywhere outside the contour increases with the number of dimensions, since there are more combinations of variables that are outside the contour. For example, in 4D, for an IFORM contour defined at a marginal exceedance probability $\alpha_m = 10^{-5}$ the total exceedance probability is approximately 100 times higher.

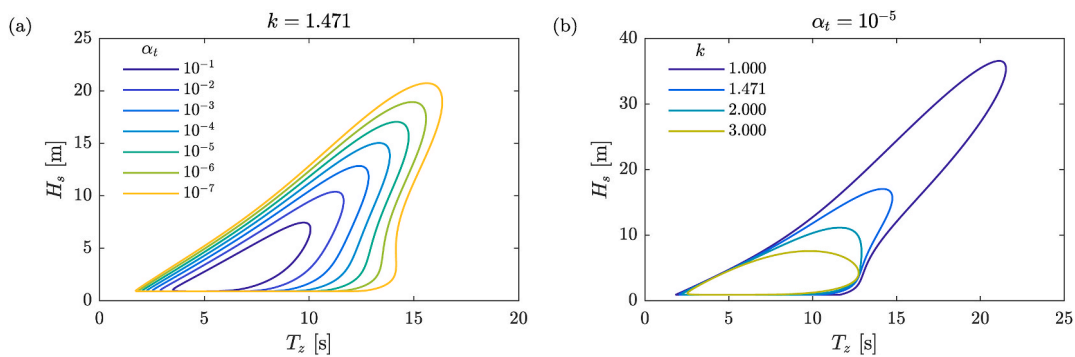

**Fig. 15.** (a) HD contours of joint density function described in (14)-(16) at various contour exceedance probabilities $\alpha_t$. (b) HD contours at exceedance probability $\alpha_t = 10^{-5}$ for same joint distribution shown in (a), but with various values of Weibull shape parameter $k$ (all other parameters held constant).

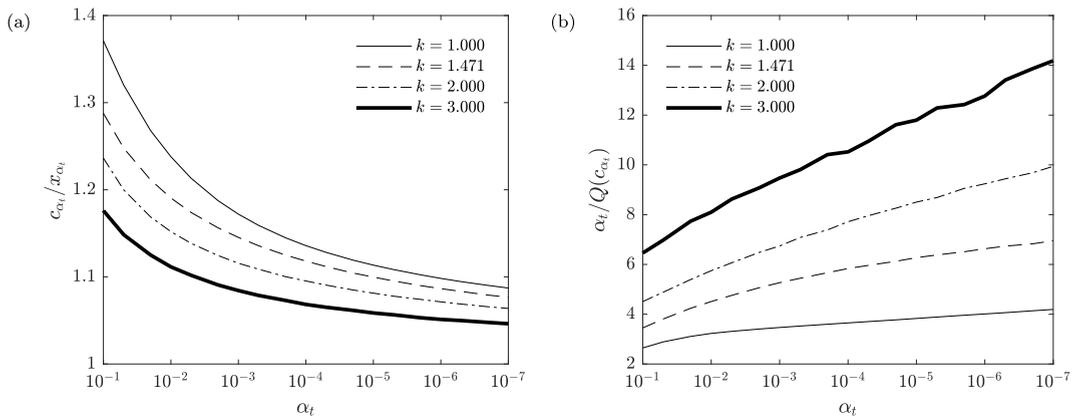

**Fig. 16.** (a) Ratio of $H_s$ return values for HD contours of joint distribution shown in previous figure. (b) Ratio of contour exceedance probability to marginal exceedance probability of highest $H_s$ value along contour.

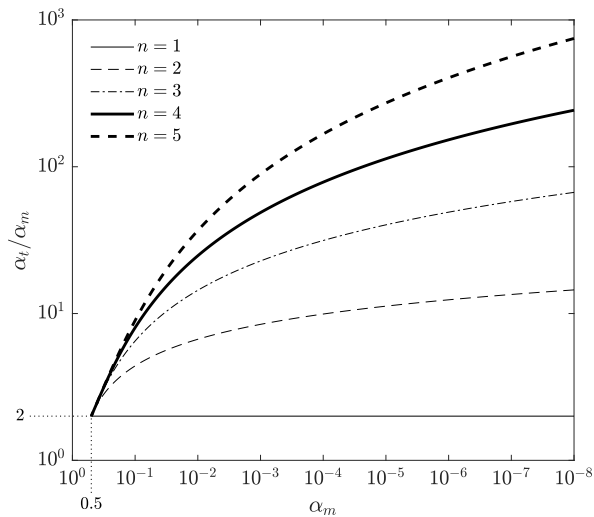

**Fig. 17.** Ratio of total exceedance probability, $\alpha_t$, to marginal exceedance probability $\alpha_m$, against $\alpha_m$ for IFORM contours defined at exceedance probability $\alpha_m$ in $n = 1, \ldots, 5$ dimensions.

*4.2. DS contours*

The probability that an observation falls outside an DS contour was investigated using numerical simulation. The joint distribution of $H_s$ and $T_z$ used in Section 3.3 was used to simulate a sample of $10^8$ observations. DS contours were calculated with marginal exceedance probabilities in the range $\alpha_m \in [10^{-6}, 10^{-1}]$. For each marginal exceedance probability, the number of points falling outside the contour was counted to obtain an empirical estimate of $\alpha_t$. Fig. 18 shows the ratio $\alpha_t/\alpha_m$ against $\alpha_m$ for the DS contour and an IFORM contour. The ratio of total to marginal exceedance probability for the DS contour is almost identical to that for an IFORM contour for exceedance probabilities greater than $10^{-3}$, but the ratio for the DS contour is slightly lower than that for the IFORM contour at lower exceedance probabilities.

DS contours can be subject to significant sampling uncertainty if the ratio of the sample size to the return period of the contour is not sufficiently large. Techniques such as importance sampling can be used to reduce the need for generating very large samples (see Refs. [40]). However, in this case the sample size of $10^8$ was sufficiently large that the ratio $\alpha_t/\alpha_m$ showed negligible variation over repeated simulations for $\alpha_m$ up to $10^{-6}$.

The IFORM contour can be thought of as a special case of the DS contour, when the data are independent normal variables. In this special case, the ratio $\alpha_t/\alpha_m$ would be identical for both DS and IFORM contours at all exceedance probabilities. In general, we would expect DS and IFORM contours to be very closely matched when the isodensity contours of the joint distribution are convex. The difference in the ratio at low exceedance probabilities shown in Fig. 18 is likely to be related to the shape of the joint distribution, which becomes concave at lower exceedance probabilities (see Fig. 15(a)). As the DS contour is convex and the IFORM contour is concave, this may cause some differences in the ratio $\alpha_t/\alpha_m$. In general, the relationship between the total probability outside the contour and the marginal exceedance probability is likely to be dependent on the shape of the joint distribution. However, as DS and IFORM contours are both defined in terms of marginal exceedances, the ratio $\alpha_t/\alpha_m$ for DS contours is likely to be similar to that for IFORM contours in higher dimensions as well.

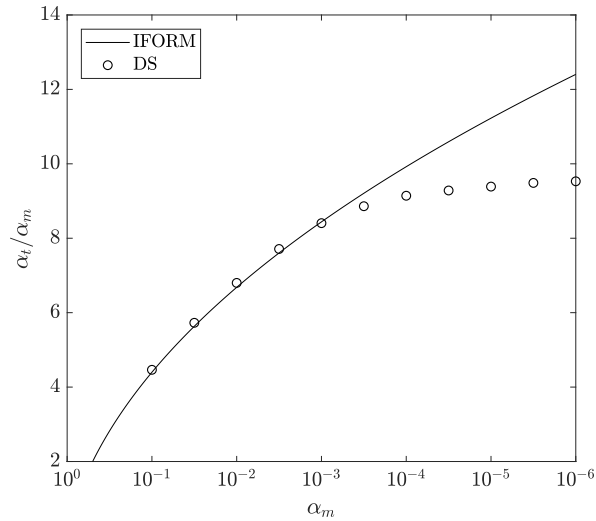

**Fig. 18.** Comparison of ratio of total exceedance probability, $\alpha_t$, to marginal exceedance probability $\alpha_m$, against $\alpha_m$ for DS and IFORM contours in 2 dimensions. Results for the DS contour are based on the joint distribution given in (14).

## 5. Implications for the design process

The previous sections showed that IFORM, DS, ISORM and HD contours that are calculated using the same α-value can lead to significantly different design conditions. Further, the marginal return periods of the maximum values of each variable along different kinds of contours can easily deviate by a factor of more than 10. Thus, the type of contour used in the design process of a marine structure can have an important influence on the assumed environmental loads and consequently on the structure that is being designed. As, by definition, ISORM and HD contours are more conservative than IFORM and DS contours, it seems sensible to use these conservative contours only when they are required, that is, when the structure of interest has a non-convex failure region. In a design project one does not know the exact shape of the failure region beforehand, however, one can often anticipate its approximate shape. In the following we consider three simple examples, whose response characteristics are representative of wider classes of structural problems. In the first example we consider the response of a single degree of freedom system to sea states, where the use of IFORM or DS contours is more appropriate. In the second example, we consider a structure, which responds strongly at two frequencies such that the failure region is not convex and IFORM and DS contours are not conservative for estimating extreme responses. Finally, in the third example we consider the specification of directional design conditions, where the use of ISORM or HD contours is more appropriate. To simplify the examples, it is assumed that the structural responses are deterministic so that it is not necessary to use inflation factors for the contours.

### 5.1. Example 1: response of a single degree of freedom system to sea states

Consider the design process of a marine structure whose behaviour can be approximated as a single degree of freedom vibration system (see, for example, [3, Chapter 2]). This could be, for example, a vibrating tower [3, pp. 23] or a floating system [41]. Assume that a candidate design needs to be evaluated to check whether it fulfills a given reliability target. Suppose that failure is only allowed to occur with a return period of 50 years (or more seldom) such that a 50-year environmental contour is constructed to derive a set of design conditions.

We assume that the candidate design responds with the following deterministic response function, proposed in Refs. [2]:

$$r(h_s, t_p) = a \frac{h_s}{1 + b(t_p - t_{p0})^2} \tag{18}$$

where $a = 2$, $b = 0.007$ and $t_{p0} = 30$ s. The response function peaks at a particular period $t_{p0}$, which could represent the eigenfrequency of the system. In this example, the peak response occurs at a period, which is higher than the wave periods of typical sea states, which could represent the heave motions of a spar or surge motions of a tension leg platform [42]. As the response function peaks at a single $t_p$ value, any response surface and consequently also the failure region will be convex.

In this example, we assume that the joint distribution of $h_s$ and $t_p$ at the location of the structure is described by the model presented in Section 4.2, with a fixed relationship between $t_p$ and $t_z$ given by $t_p = 1.2796 t_z$. The 50-year IFORM, ISORM, DS and HD contours for this environmental model are shown in Fig. 19 and the position of the highest response along each contour is marked with a cross. The highest response along each contour is listed in Table 1 and ranges from 14.5 (IFORM contour) to 18.2 (ISORM contour).

The true long-term distribution can be calculated using the expression

$$F_R(x) = \iint\limits_{r(h_s,t_p) \leq x} f_{H_s,T_p}(h_s,t_p) \mathrm{d}t_p \mathrm{d}h_s. \tag{19}$$

The true 50-year response for this example was calculated as 14.5. Thus, the computed IFORM and DS responses deviate less than 2% from the true 50-year response, whereas the responses from the HD and ISORM contours are 18% and 26% higher respectively than the true 50-year response.

Suppose that the structure has been optimized with respect to the design conditions that were derived from the 50-year DS contour such that a response that exceeds 15 represents failure. The resulting convex failure region would almost touch the IFORM and DS contours, but the failure region would overlap with the ISORM and HD contours (Fig. 19). Consequently, if ISORM or HD contours were used, the candidate design would need to be changed, either to reduce the response or increase the capacity.

In this example, the failure region is convex and consequently the probability of failure is less than the DS contour's probability of exceedance, $\alpha_M$. Consequently, using the DS or IFORM contour would be a better choice than using an HD or ISORM contour as otherwise unnecessary conservatism would be introduced.

If the structure has multiple eigenfrequencies, however, the failure region will not be convex and it is not clear whether the exceedance probability of an DS or an IFORM contour, $\alpha_m$, is greater than the structure's probability of failure. In these cases, further analysis is required to ensure that the probability of failure is below the required value. This problem is considered further in the next section.

*5.2. Example 2: bimodal response to sea states*

Suppose that the structure of interest is not a single degree of freedom system, but that it responds strongly at two distinct frequencies. These two frequencies could be the effect of two eigenfrequencies. We assume that the structure responds with a function that is the sum of two response functions of the form given in (18):

$$r(h_s, t_p) = a_1 \frac{h_s}{1 + b_1(t_p - t_{e1})^2} + a_2 \frac{h_s}{1 + b_2(t_p - t_{e2})^2}, \tag{20}$$

where $a_1 = 4$, $a_2 = 1.1$, $b_1 = 0.1$, $b_2 = 0.05$, $t_{e1} = 25$ s and $t_{e2} = 12.5$ s. These parameter values give a failure region that is concave in the area around the highest $H_s$ values on the contours. It will be shown below that in this case, the assumption of a linear failure surface in the IFORM and DS contours is non-conservative. Not all bimodal responses result in non-conservative estimates from IFORM and DS contours, but such a case is used here to illustrate that special care is required for structures with multi-modal responses.

Suppose that the structure considered is subject to the same environment as the structure in Example 1 and that we are interested in the 50-year response. As the environment is the same, the environmental contours are the same too. Based on these contours, the 50-year response is estimated as the largest response along the contour. The estimated responses, listed in Table 2, range from 16.57 (IFORM contour) to 20.47 (ISORM contour). The true 50-year response was calculated using (19) as 17.00. Thus, in this example, response estimators from IFORM and DS contours are not conservative.

Fig. 20 shows the environmental contours together with a failure surface - assuming that failure occurs when the true 50-year response, $r = 17.00$, is exceeded. The IFORM and DS contours do not touch this failure surface, while the ISORM and HD contours overlap with the failure region. Consequently, in this case, a designer who used an IFORM or DS contour could see potential to optimize the structure by reducing the structure's capacity, for example to $r_{cap} = 16.7$. Such a response capacity, however, would lead to a probability of failure greater than the target probability of failure. A structure with a response capacity that is fully optimized to the IFORM 50-year response, $r_{cap} = 16.57$, would have a failure probability that is 64% higher than the target probability of failure. Using the responses estimated using ISORM or HD contours, however, would also have unintended consequences: A design with a capacity of $r_{cap} = 17$, which would lead to the intended target probability of failure, would be dismissed and instead a more conservative design would be created.

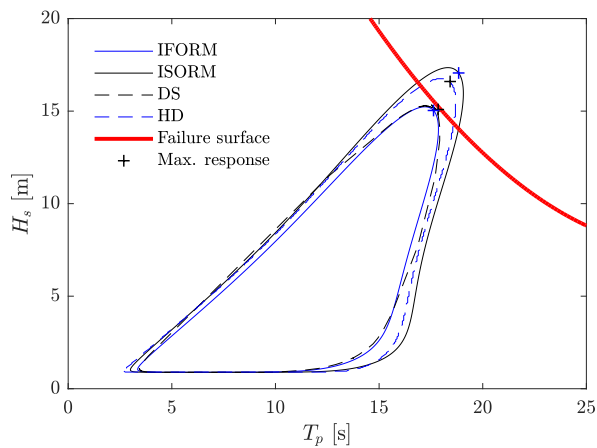

**Fig. 19.** Design of a single degree of freedom system with a sea state environmental contour. The assumed response function peaks at the eigenfrequency, $t_{p0} = 30$ s. The failure surface would be acceptable if an IFORM or DS contour was used, but not acceptable if an HD or ISORM contour were used.

**Table 1**
Structural responses from different contour methods for a single degree of freedom system.

| Method | 50-year response | Sea state causing 50-year response ($h_s$, $t_p$) |
| --- | --- | --- |
| IFORM contour | 14.51 | 15.0 m, 17.6 s |
| DS contour | 14.55 | 15.0 m, 17.7 s |
| ISORM contour | 18.23 | 17.1 m, 18.8 s |
| HD contour | 17.13 | 16.6 m, 18.4 s |
| all sea states approach | 14.53 | – |

*5.3. Example 3: directional design conditions*

In this example we consider the specification of directional design conditions for the design or assessment of a marine structure. In many locations, the severity of wave conditions exhibits a dependence on wave direction. Specifying the design wave height as a function of direction can allow the optimisation of an asymmetrical structure. Directional return values are often estimated in discrete directional sectors. However, it is becoming more common to estimate extreme value models where the model parameters vary smoothly with direction (e.g. Refs. [43,44]). In these cases, environmental contours can be used to define directional design conditions.

In this example, we use a simple description of the joint distribution of $H_s$ and wave direction, $\Theta$, based on the joint distribution of wind speed and direction given in Ref. [45] and studied further in Ref. [46]. Its mathematical description is given in Appendix A.

When constructing contours for directional distributions it is important to be clear about what coordinate system is used. If IFORM, ISORM or DS contours are constructed in the $H_s$-$\Theta$ coordinate system, then the contours will have upper and lower bounds for $\Theta$ for any directional distribution, which does not make sense from the perspective of structural reliability. Whether HD contours have upper and lower bounds on $\Theta$ depends on the shape of the joint PDF. The construction of IFORM contours based on applying the Rosenblatt transformation in the form $\Phi(U_1) = F_\Theta(\theta)$, $\Phi(U_2) = F_{H_s|\Theta}(h_s|\theta)$, was discussed in Refs. [45–47]. The IFORM contours presented in these works exhibit a discontinuity in the transition between 0 and $2\pi$, due to the upper and lower bounds for $\theta$.

To address this issue, Vanem et al. [46] proposed deriving contours in terms of the *x* and *y* components of significant wave height, defined as

$$h_x = h_s \cos(\theta), \quad h_y = h_s \sin(\theta). \tag{21}$$

The joint density function of $h_x$ and $h_y$ in Cartesian coordinates can be written as

$$f_{H_x,H_y}(h_x, h_y) = f_{\Theta,H_s}(\theta, h_s) \left| \frac{\partial(h_s \cos(\theta), h_s \sin(\theta))}{\partial(h_s, \theta)} \right|^{-1} = \frac{f_{\Theta,H_s}(\theta, h_s)}{h_s}. \tag{22}$$

Note that since the Weibull location parameter, $\gamma$, is greater than zero in this example, the transformed density function is non-singular at the origin since $f_{\Theta,H_s}(\theta, 0) = 0$. In the following, all the contours considered are constructed in $H_x$-$H_y$ space. This means that DS contours are based on the assumption that the failure surface is linear in this space. For IFORM contours, the assumed failure surface is not necessarily linear in $H_x$-$H_y$ space, due to the use of the Rosenblatt transformation. However, this is a common feature of the IFORM method for any parameter space, as discussed in Refs. [7,11,40].

The 1-year IFORM, ISORM, DS and HD environmental contours are shown in Fig. 21(a) together with the 1-year omnidirectional return value (assuming that observations are at 3 h intervals as before). As the isodensity contours of the joint distribution are close to convex in $(h_x, h_y)$ coordinates, the DS and IFORM contours are in close agreement, as are the HD and ISORM contours. The HD and ISORM contours give maxim values of $H_s$ approximately 1 m larger than the DS and IFORM contours. The IFORM and ISORM contours exhibit a small ridge at around 90°. This is a result of the use of the Rosenblatt transformation. The location of the ridge is dependent on the order of variables used in the transformation, as illustrated in Fig. 21(b).

**Table 2**
Structural responses from different contour methods for a structure with a bimodal response function.

| Method | 50-year response | Sea state causing 50-year response ($h_s$, $t_p$) |
| --- | --- | --- |
| IFORM contour | 16.57 | 15.2 m, 17.4 s |
| DS contour | 16.63 | 15.2 m, 17.5 s |
| ISORM contour | 20.47 | 17.0 m, 18.9 s |
| HD contour | 19.10 | 16.6 m, 18.4 s |
| all sea states approach | 17.00 | – |

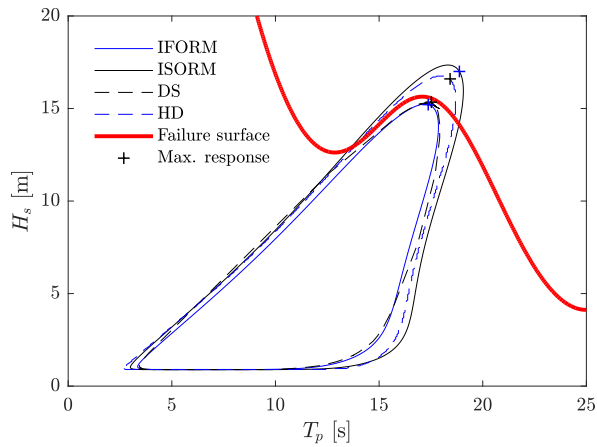

**Fig. 20.** Sea state environmental contours for the design of a structure with a bimodal response. The assumed response function peaks at $t_{e1} = 25$ s and at $t_{e2} = 12.5$ s such that the failure region is not convex. The failure surface shown is based on the capacity that causes the failure probability to become exactly α ($r_{cap} = 17$). If designers were to use the IFORM or DS contour they might design a structure with $r_{cap} < 17$, while designers using the ISORM or HD contour would be required to increase the structure's capacity above $r_{cap} = 17$.

Note that marginal probabilities in Cartesian $H_x - H_y$ space do not correspond to marginal probabilities of $H_s$. The set of points with $H_s \leq r$ is contained in the set of points with $H_x \leq r$ (see Fig. 22). Since this is true under any rotation of the axes, the maximum radius of a DS contour constructed in $H_x - H_y$ space will be less than the omnidirectional return value. In this case the 1-year omnidirectional return value is 8.65 m and the maximum $H_s$ on the DS contour is 8.58 m. The maximum $H_s$ on the IFORM contour is slightly higher, at 8.59 m, due to the effects of the Rosenblatt transformation, discussed in Section 3.1.

Suppose that a structure has a deterministic omnidirectional response and is designed so that it fails if the wave height exceeds the maximum value along a $T$-year IFORM or DS contour. Since the $T$-year omnidirectional return value of $H_s$ is greater than the maximum value of $H_s$ along the contour, the structure will have a higher failure probability than intended. Now suppose that the omnidirectional structure is designed using the omnidirectional return value at exceedance probability α, denoted $h_\alpha$, so that the failure surface is located at $h_\alpha$ and is independent of direction. In this case the probability of failure will be equal to α, by definition, since $h_\alpha$ is the value of $H_s$ that is exceeded with probability α, independent of direction. If the structure is asymmetric, and designed so that the failure surface is located at a value greater than or equal to $h_\alpha$ in each direction, then the probability of failure will be less than or equal to α. So the use of omnidirectional criteria gives the target failure probability, but may lead to a less efficient design as the potential to optimize the design with respect to direction is not exploited.

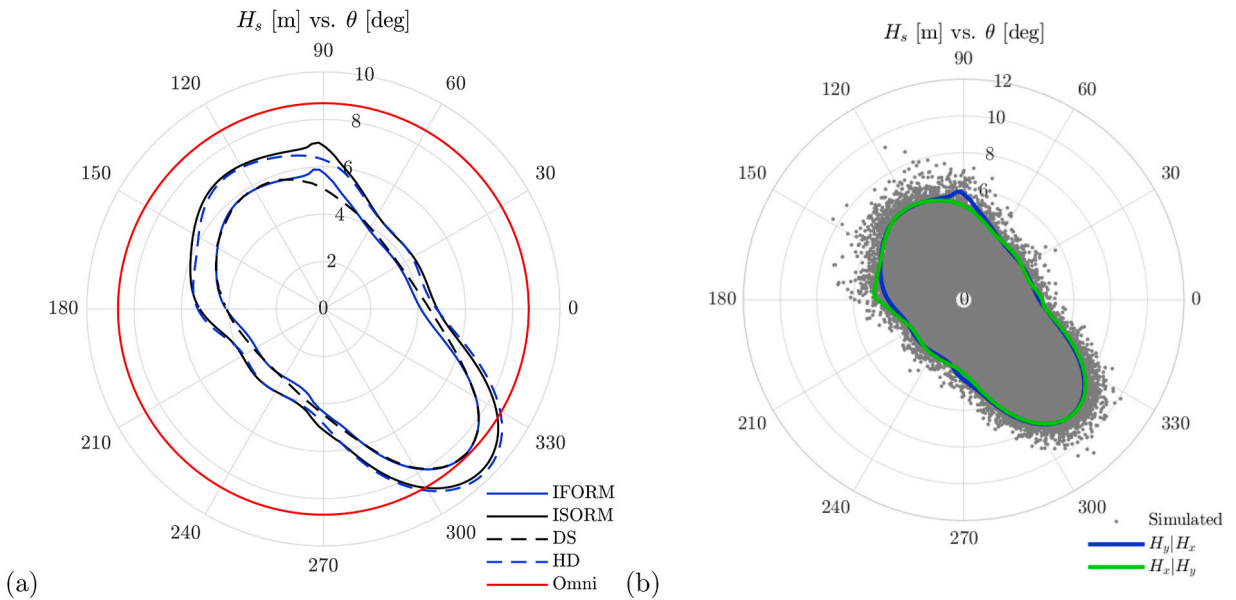

**Fig. 21.** (a) 1-year environmental contours for joint distribution of $H_s$ and wave direction and 1-year omnidirectional return value. IFORM and ISORM contours based on transformation in terms of conditional distribution $H_y|H_x$. (b) Effect of the order of variables used in the Rosenblatt transformation used to construct the IFORM contour.

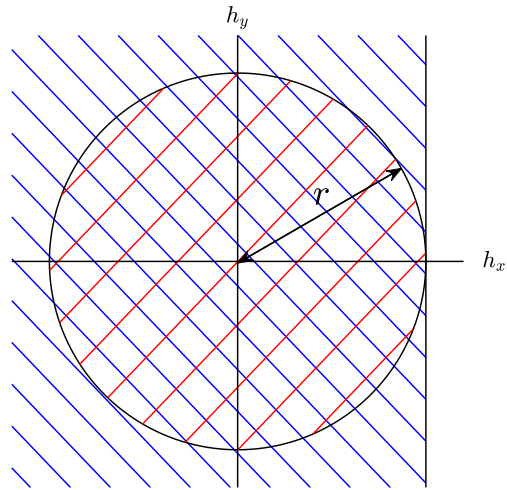

**Fig. 22.** The set of points with $H_s \leq r$ (red hatched area) is contained in the set of points with $H_x \leq r$ (blue hatched area). The maximum $H_s$ on a $T$-year DS contour is therefore smaller than the $T$-year omnidirectional $H_s$ return value. For the red hatched area to contain the same probability as the blue hatched area, the radius $r$ must be greater than the bound on $h_x$. (For interpretation of the references to colour in this figure legend, the reader is referred to the Web version of this article.)

Now suppose that an asymmetric structure is optimized based on directional extreme conditions from an environmental contour. For example, the stiffness of a fixed structure in a particular direction could be optimized with respect to direction or similarly the mooring response of a floating structure could be optimized to allow larger responses in directions where the wave conditions are less severe.

Consider a hypothetical marine structure that is fully optimized to directional values of $H_s$, such that failure occurs immediately if the directional value of $H_s$ is exceeded. In such a case we can see from the results in Sections 4.1 and 4.2 that if the directional values of $H_s$ are specified based on IFORM or DS contours then the failure probability will be approximately 10 times higher than the exceedance probability of the contour. So for this application the use of IFORM or DS contours is not conservative. A similar argument was made by Forristall [48]. He noted that if a structure is designed based on directional return values in discrete sectors, each at exceedance probability α, then the failure probability of the structure is greater than α. The argument made here is analogous, although a distinction should be made between directional IFORM and DS contours and directional return values. The former are continuous with direction, but their interpretation in terms of exceedance values of $H_s$ is less clear. In contrast, directional return values have a clearer interpretation, but are discontinuous with direction and are dependent on the sector width used (see Appendix B).

ISORM and HD contours at exceedance probability α both have a total exceedance probability of α. Therefore, if the hypothetical, fully optimized structure is designed using directional criteria derived from ISORM or HD contours then the failure probability will be equal to the contour exceedance probability (under the assumption of a deterministic response dependent on $H_s$ and Θ only).

The ISORM and HD contours as well as the omnidirectional return value contour all enclose a region of parameter space containing total probability $1 - α$. From Fig. 21 it is evident that the ISORM and HD contours offer the potential for considerable directional optimisation compared to the omnidirectional contour, but at the cost of requiring a greater capacity in the sector between 300° and 330°. Feld et al. [49] note that there are infinitely many ways to define directional criteria which achieve the target reliability for the structure. The use of directional return values specified in discrete sectors has the disadvantage that the return periods used in each sector are dependent on the sector width, with the use of smaller sectors requiring larger return periods in order to maintain the desired reliability (see Appendix B). This is mainly a problem for communication of results, rather than a problem for the optimisation of a structure. Since it may not be immediately clear that directional return values at a given return period are dependent on the sector width, there has been some resistance to the use of directional design criteria with different return periods to the omnidirectional design value. The use of ISORM or HD contours in this context provides a set of directional design criteria with a definition that is easier to interpret. As discussed above, there is no fixed relation between the largest values of $H_s$ on environmental contours defined in $h_x$-$h_y$ parameter space and return values of $H_s$. However, since directional return values are dependent on the width of directional sector used, it could be argued that directional return values are somewhat ambiguous quantities anyway. One way to interpret the omnidirectional return value is that it is the contour with the smallest radius containing probability $1 - α$. Similarly, the HD contour defines the smallest area containing probability $1 - α$. The interpretation of the ISORM contour is less clear due to the use of the Rosenblatt transformation.

In reality no structure will be fully optimized in each direction. Instead, there could be a single direction where exceeding the directional values leads to failure, but in other directions a non-zero margin exists until failure occurs. In such a case the probability of failure of a structure that is designed based on design conditions from HD and ISORM contours will be smaller than the target failure probability. Consider the simple example that the structure is designed to have a smaller response in the $x'$-direction than the $y'$-direction (where $x'$ and $y'$ represent a local coordinate system of the structure). In this example we consider the following simple response function

$$r(h_{x'}, h_{y'}) = \sqrt{ah_{x'}^2 + bh_{y'}^2}, \qquad (23)$$

where $a = 1.4$ and $b = 5$. Or expressing this in the global coordinates, $x, y$, which are rotated by an angle $\varphi$ to the local coordinate system:

$$r(h_x, h_y) = \sqrt{a(h_x\cos\varphi + h_y\sin\varphi)^2 + b(-h_x\sin\varphi + h_y\cos\varphi)^2}. \qquad (24)$$

Based on the distribution of $H_s$ and $\Theta$ we are considering here, suppose that the structure is designed such that its response per unit wave height is smallest in the direction where the highest waves occur, such that $\varphi = 315°$. The 1-year response of this structure was calculated using the all sea states approach and using the four environmental contour methods. The results are listed in Table 3. The response surface corresponding to (24) is shown in Fig. 23, where the response value corresponds to the 1-year response estimated using the all sea states approach.

As expected, the IFORM and DS contours underestimate the 1-year response, since the methods are based on the assumption that the failure region is convex in $H_x$-$H_y$ space. Conversely, the ISORM and HD contours overestimate the 1-year response. The underestimation is approximately 3–5% for the IFORM and DS contours and the overestimation is approximately 7–16% for the ISORM and HD contours.

Suppose that the structure is designed such that its capacity is exactly the 1-year response estimated by a particular contour method such that, for example, in the case that an IFORM contour is used the capacity is 10.33, but if an ISORM contour is used the capacity is 12.37. In this case the probability of failure for IFORM and DS contours are roughly double the target probability of failure and the probability of failure for ISORM and HD contours are lower than the target probability of failure (roughly 20 and 4 times lower, respectively). So in this example, the use of IFORM or DS contours would be non-conservative, whilst using ISORM or HD contours would achieve the target reliability.

The directional response function given in (24) is not intended to represent a particular structure, but is intended to illustrate the differences between the various contour methods considered. The response of a particular structure will fall somewhere between the two limiting cases considered above: (i) a fully optimized structure whose failure surface coincides with the environmental contour that was used to design it; and (ii) an omnidirectional structure whose capacity is designed to withstand the highest response along the contour. Table 4 lists the failure probabilities of these two structures when designed using each contour method. Interpreting these cases as the boundaries of the set of reasonable direction-sensitive structures, for the joint distribution of $H_s$ and $\Theta$ considered here, a structure designed using an IFORM or DS contour will have a probability of failure between ca. 1.1 and 9.2 $\alpha$ and a structure designed using an ISORM or HD contour will have a probability of failure between ca. 0.07 and $\alpha$.

## 6. Summary and conclusions

The differences between environmental contours defined in terms of marginal and total exceedance probabilities have been explored. A summary of the properties of the four contour methods considered is given in Table 5. The ratio between the marginal exceedance probability of the largest value of each marginal variable along a contour and the total exceedance probability of the contour depends on the contour exceedance probability, the number of variables and the shape of the joint distribution. For IFORM and ISORM contours the total exceedance probability is independent of the joint distribution, but the marginal exceedance probability of the largest values of each variable along the contour depends on the order in which variables are used in the Rosenblatt transformation. The marginal return period of the largest marginal value along an ISORM contour in 2D, is approximately 10 times higher than the contour return period, whereas in 4D the marginal return period of the largest marginal value along an ISORM contour can be more than 200 times higher.

The relationship between marginal and total exceedance probabilities for direct sampling contours is dependent on the shape of the joint distribution of variables. However, in the case that the joint distribution has convex isodensity contours, the relationship between marginal and total exceedance probabilities is approximately equal to that for IFORM contours.

**Table 3**
Structural responses from different contour methods for a system whose response depends on wave direction, with response given by (24). $P_f$ is the probability of failure of a structure that is designed to have a capacity of the 1-year response.

| Method | 1-year response | Sea state causing 1-year response ($h_s$, θ) | $P_f$ |
| --- | --- | --- | --- |
| IFORM contour | 10.33 | 5.8 m, 91° | 1.80 α |
| DS contour | 10.20 | 8.6 m, 311° | 2.28 α |
| ISORM contour | 12.37 | 7.0 m, 91° | 0.05 α |
| HD contour | 11.49 | 9.7 m, 311° | 0.23 α |
| All sea states approach | 10.7 | – | α |

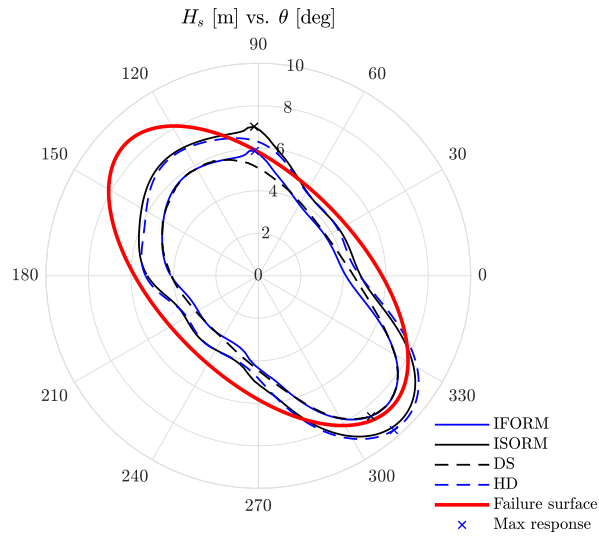

**Fig. 23.** 1-year environmental contours for joint distribution of $H_s$ and direction, with failure surface for structure with response given by (24). Note that the IFORM and ISORM contours have their highest response at the small ridge around 90°, which is an effect of the order of variables used in the Rosenblatt transformation (Fig. 21 b).

**Table 4**
Structural responses from different contour methods for the two limiting cases of directionally optimized structures: A hypothetical structure that is fully optimized to wave direction and a structure whose response is independent of wave direction (computed by setting $a = b$ in (24)). $P_f$ is the probability of failure of a structure that is designed to have a capacity of the environmental contour method's estimated 1-year response.

| Method | $P_f$ of fully optimized structure | $P_f$ of omni-directional structure |
| --- | --- | --- |
| IFORM contour | 9.16 α (see (17)) | 1.16 α |
| DS contour | ca. 9 α (similar to IFORM) | 1.11 α |
| ISORM contour | α | 0.11 α |
| HD contour | α | 0.07 α |

**Table 5**
Summary of properties of environmental contours.

| Property | IFORM | DS | ISORM | HD |
| --- | --- | --- | --- | --- |
| Defining exceedance probability | Marginal | Marginal | Total | Total |
| Approximation to failure region | Linear boundary in $U$-space | Linear boundary in $X$-space | Surrounds contour | Surrounds contour |
| Contour shape | Concave or convex | Convex | Concave or convex | Concave or convex |
| Contour is uniquely defined | No[a] | Yes | No[a] | Yes |
| Marginal exceedance probabilities independent of joint distribution | No[b] | Yes | No[b] | No |
| Total exceedance probability independent of joint distribution | Yes | No | Yes | Yes |

[a] Contour depends on order of variables in Rosenblatt transformation.
[b] Marginal exceedance probability is only fixed for conditioning variable in Rosenblatt transformation.

For highest density contours, the marginal exceedance probability of the largest marginal value along the contour is not in fixed relationship with the total exceedance probability, but depends on the shape of the joint distribution of variables. In general we have $c_\alpha > x_\alpha$ for HD contours in two or more dimensions.

The effect this has on structural reliability calculations is problem specific, with the impact depending on both the joint distribution of variables and the response function. In particular, the appropriate choice of contour is dependent on the shape of the failure surface. In the example for a simple one-degree-of-freedom structure, the failure region was convex and the use of IFORM or DS contours was appropriate. In this case, the 50-year IFORM and DS contours gave an estimate of the 50-year response within 2% of the true 50-year response, whereas the HD and ISORM contours gave estimates that were 18% and 26% higher respectively. Clearly, this can have significant implications for the design of a structure. In the second example, where the structure had a bimodal response, the estimates of the 50-year response from the IFORM and DS contours was no longer conservative, as the structure's failure region was concave in the area around the largest values of $H_s$ on the contours. Whilst IFORM and DS contours may be conservative for some structures with multi-modal responses, the example illustrates that special care is needed for these cases.

The third example considered the specification of directional design conditions. It was argued that it does not make sense to define contours using $H_s$ and $\Theta$ as orthogonal coordinates on Cartesian axes, since this results in contours with upper and lower bounds on direction. Instead, it makes more sense to define directional contours in terms of the $x$ and $y$ components of $H_s$. In this coordinate system the failure region of most marine structures will not be convex. In this case, the use of DS or IFORM contours results in non-conservative design conditions, whereas HD or ISORM contours maintain the intended reliability.

The results presented in this work emphasise the importance of having some understanding of the shape of the failure surface of a structure in the coordinate system that the contours are constructed in, so that the appropriate choice of environmental contour can be made.

**Data availability**

The analysis of the examples presented in Section 5 can be reproduced by running the MATLAB files Example1and2.m and Example3.m that are available at the GitHub repository https://github.com/ahaselsteiner/2020-note-on-contours. Matlab implementations of the IFORM, ISORM, DS and HD methods are available in the software package compute-hdc that is available at https://github.com/ahaselsteiner/compute-hdc.

**Declaration of competing interest**

The authors declare that they have no known competing financial interests or personal relationships that could have appeared to influence the work reported in this paper.

**Acknowledgements**

We thank the reviewers for suggestions and comments on an earlier version of the manuscript, which helped to clarify numerous aspects of the work. EM was funded under the UK Engineering and Physical Sciences Research Council (EPSRC) grant EP/R007519/1 and as part of the Tidal Stream Industry Energiser Project (TIGER), which has received funding from the European Union's INTERREG V A France (Channel) England Research and Innovation Programme, which is co-financed by the European Regional Development Fund (ERDF).

**Appendix A. Directional wave distribution in Example 3**

The joint density function of wave height and direction considered in Example 3 is written as a hierarchical model:

$$f_{\Theta,H_s}(\theta,h_s) = f_\Theta(\theta)f_{H_s|\Theta}(h_s|\theta). \quad (A.1)$$

The marginal distribution of wave direction is expressed as a mixture of von Mises distributions,

$$f_\Theta(\theta) = \sum_{i=1}^{n} w_i f_i(\theta), \quad (A.2)$$

where $w_i \in [0,1]$ are weights and $\sum_{i=1}^{n} w_i = 1$. The PDF of the von Mises distribution is

$$f_i(\theta) = \frac{\exp(\kappa_i \cos(\theta - \mu_i))}{2\pi I_0(\kappa_i)}, \quad (A.3)$$

where $I_0$ is the zero-order modified Bessel function of the first kind, $\kappa_i$ is a concentration factor and $\mu_i$ is the location parameter. The model for $H_s$ conditional on direction is a three-parameter Weibull distribution (13), with parameter dependence on direction given in terms of a Fourier series:

$$\lambda(\theta) = a_0 + \sum_{j=1}^{m} a_j \cos(j\theta) + b_j \sin(j\theta), \quad (A.4)$$

$$k(\theta) = c_0 + \sum_{j=1}^{m} c_j \cos(j\theta) + d_j \sin(j\theta). \quad (A.5)$$

The model presented in Refs. [45] uses $m = 8$ harmonics for the Fourier series. In this example we assume the Weibull location parameter is constant at $\gamma = 0.5$. The other distribution parameters are defined in Tables A1 and A.2.

Table A 1
Parameters for distribution of wave direction.

| j | $w_j$ | $\mu_j$ | $\kappa_j$ |
|---|-------|---------|------------|
| 1 | 0.21  | 2.10    | 0.74       |
| 2 | 0.79  | 5.54    | 13.11      |

Table A 2
Fourier coefficients for distribution of $H_s$ conditional on wave direction.

| j | $a_j$  | $b_j$  | $c_j$   | $d_j$  |
|---|--------|--------|---------|--------|
| 0 | 1.875  |        | 1.910   |        |
| 1 | 0.345  | −0.140 | 0.240   | −0.130 |
| 2 | −0.210 | −0.820 | −0.080  | −0.170 |
| 3 | −0.160 | −0.200 | −0.010  | −0.030 |
| 4 | −0.265 | 0.095  | −0.110  | 0.030  |
| 5 | −0.090 | 0.110  | 0.0004  | 0.003  |
| 6 | 0.070  | 0.070  | 0.060   | 0.020  |
| 7 | 0.030  | 0.020  | 0.060   | 0.020  |
| 8 | 0.030  | −0.015 | 0.004   | −0.010 |

**Appendix B. Directional return values**

Forrsitall's argument about the use of directional design values [48] can be summarised as follows. Suppose we have $M$ independent observations of $H_s$ per year with CDF $F_{H_s}$. The $T$-year omnidirectional return value, $H_T$, is defined as the solution of

$$1 - F_{H_s}(H_T) = \frac{1}{MT}. \quad (B.1)$$

Suppose also that the direction of each observation is random and uniformly distributed in $[0, 360°]$ and that the distribution of $H_s$ is independent of direction. The $T$-year return value in a directional sector of width $d$ degrees, denoted $H_T^d$, is the solution of

$$1 - F_{H_s}(H_T^d) = \frac{1}{mT}, \quad (B.2)$$

where $m = M d/360$. Since $m < M$ for $d < 360$, the exceedance probability of $H_T^d$ is greater than the exceedance probability of $H_T$ and hence $H_T^d < H_T$. So when the distribution of $H_s$ and storm occurrence rate are constant with direction, the $T$-year return values in a directional sector is less than the $T$-year omnidirectional return value and $T$-year directional return value decreases as the sector width decreases.

Clearly, in this situation, a structure designed using $T$-year directional return values will have a lower reliability than a structure designed using the $T$-year omnidirectional return value. Therefore, to ensure the overall reliability remains the same when using directional criteria, the directional return periods must be increased so that the probability that the design criteria are exceeded in any directional sector (i.e. the sum of exceedance probabilities in each sector) is equal to the target probability of failure.

This argument extends to the case where both the distribution of $H_s$ and the occurrence rate varies with direction. For further details see Ref. [48,49].